\documentclass[prd,amsmath,superscriptaddress, twocolumn,nofootinbib]{revtex4-2}
\pdfoutput=1

\usepackage[utf8]{inputenc}
\usepackage{graphicx}
\usepackage{float}
\usepackage{epsf,latexsym,bbm,euscript}
\usepackage{amssymb,amsmath}
\usepackage{mathtools} 
\usepackage{times}
\usepackage{soul,xcolor}
\usepackage{mathtools}
\usepackage{mathrsfs}
\usepackage{array}
\usepackage{booktabs}
\usepackage{enumitem}
\usepackage{tensor}
\usepackage[dvipsnames]{xcolor}

\usepackage[caption=false]{subfig}
\newcommand{\RN}[1]{\uppercase\expandafter{\romannumeral #1\relax}}

\usepackage{environ}
\NewEnviron{thincases}{\scalebox{0.5}[1]{$\displaystyle
		\left\{\scalebox{2}[1]{\setlength{\arraycolsep}{6pt}
			$\displaystyle\begin{array}{ll}
				\BODY
			\end{array}$}\right.$}}
			\NewEnviron{thinpmatrix}{\scalebox{0.5}[1]{$\displaystyle
	\left(\scalebox{2}[1]{$\displaystyle
		\begin{matrix}
			\BODY
		\end{matrix}$}
	\right)$}}
	
	\makeatletter
	\g@addto@macro\bfseries{\boldmath}
	\makeatother
	
	\makeatletter
	\newcommand*{\defeq}{\mathrel{\rlap{%
		\raisebox{0.3ex}{$\m@th\cdot$}}%
	\raisebox{-0.3ex}{$\m@th\cdot$}}%
=}
\newcommand*{\eqdef}{=\mathrel{\rlap{%
		\raisebox{0.3ex}{$\m@th\cdot$}}%
	\raisebox{-0.3ex}{$\m@th\cdot$}}%
	}
	\makeatother

	\usepackage{pifont}
	\newcommand{\cmark}{\text{\ding{51}}}
	\newcommand{\xmark}{\text{\ding{55}}}
	
	\usepackage{scalerel}
	\usepackage{tikz}
	\usetikzlibrary{svg.path}
	\definecolor{orcidlogocol}{HTML}{A6CE39}
	\tikzset{
orcidlogo/.pic={
	\fill[orcidlogocol] svg{M256,128c0,70.7-57.3,128-128,128C57.3,256,0,198.7,0,128C0,57.3,57.3,0,128,0C198.7,0,256,57.3,256,128z};
	\fill[white] svg{M86.3,186.2H70.9V79.1h15.4v48.4V186.2z}
	svg{M108.9,79.1h41.6c39.6,0,57,28.3,57,53.6c0,27.5-21.5,53.6-56.8,53.6h-41.8V79.1z M124.3,172.4h24.5c34.9,0,42.9-26.5,42.9-39.7c0-21.5-13.7-39.7-43.7-39.7h-23.7V172.4z}
	svg{M88.7,56.8c0,5.5-4.5,10.1-10.1,10.1c-5.6,0-10.1-4.6-10.1-10.1c0-5.6,4.5-10.1,10.1-10.1C84.2,46.7,88.7,51.3,88.7,56.8z};
}
}
\newcommand\orcidlink[1]{\href{https://orcid.org/#1}{\mbox{\scalerel*{
			\begin{tikzpicture}[yscale=-1,transform shape]
				\pic{orcidlogo};
			\end{tikzpicture}
		}{X}}}}
		
		\usepackage{url,hyperref}
		\hypersetup{colorlinks,linkcolor={blue!55!black},citecolor={red!50!black},urlcolor={blue!45!black},breaklinks=true}

		\usepackage{url,hyperref}
		\hypersetup{colorlinks,linkcolor={blue!55!black},citecolor={red!50!black},urlcolor={blue!45!black}}

		\usepackage[colorinlistoftodos,prependcaption,textsize=tiny]{todonotes}

\begin{document}

\title{Regular ultracompact objects with anti-de Sitter cores as \\polymerized vacuum solutions}
\author{Hongguang Liu\orcidlink{0000-0002-1059-045X}}
\email{liuhongguang@westlake.edu.cn}

\author{Ioannis Soranidis\orcidlink{0000-0002-8652-9874}}
\email{ioannis.soranidis@westlake.edu.cn}

\affiliation{Institute for Theoretical Sciences, and Departments of Physics and Astronomy, Westlake University, Hangzhou 310030, China}
\affiliation{Institute of Natural Sciences, Westlake Institute for Advanced Study, Hangzhou 310024, China}

\begin{abstract}
	\vspace*{0.2cm}	
	We present a systematic derivation of regular black hole solutions -- and their horizonless counterparts -- that achieve regularization via an anti-de Sitter core. These geometries emerge as polymerized vacuum solutions inspired by loop quantum gravity, constituting effective quantum gravity configurations that admit a Birkhoff-type theorem and are uniquely determined by their mass. Using an auxiliary relational dust clock, together with the absence of gravitational waves in spherical symmetry, we exploit the structural ultralocality of the system to decompose the dynamics into independent shell degrees of freedom. The dust field acts as a reference clock for deparameterization and does not source the vacuum geometries considered here. These assumptions tightly constrain the Lemaître–Tolman–Bondi shell Hamiltonian to a factorized form and the static vacuum metric function to a universal expression. We examine the possibility of a bounce and analyze how its presence is encoded, or missed, in finite-order effective truncations of the full model. The procedure for deriving the explicit physical Hamiltonian is described for a generic case before specializing to a specific model of interest. Finally, we construct a four-dimensional covariant completion of the spatially covariant Lagrangian, showing that it belongs to the class of generalized extended mimetic gravity models.	
\end{abstract}

\maketitle

\section{Introduction}\label{sec:introduction}

Do the singularities of general relativity persist in a more fundamental theory, or are they resolved? This question lies at the heart of quantum gravity research. Singularities are not exceptional but generic features of classical general relativity (GR), appearing both in cosmological contexts  -- such as the big bang -- and in the gravitational collapse of massive stars \cite{HE:73}. In 1965, Penrose demonstrated that, under certain assumptions, a spacetime containing trapped surfaces is inevitably geodesically incomplete, introducing the first “modern” singularity theorem \cite{P:65}. The existence of inextensible geodesics is commonly equated with the presence of regions of unbounded curvature, known as gravitational \textit{singularities}\footnote{While we adopt this assumption in this article, it is crucial to note that this equivalence does not always hold true \cite{W:20,SW:17, SW:17-essay}}. If taken as physical, they represent a breakdown of predictability and a limitation of our theoretical framework. It is widely believed that a quantum theory of gravity will cure these pathologies, but the precise mechanism remains largely unknown outside specific models.

The phenomenological outcomes of singularity regularization have a long history of investigation. Best known among them are the Bardeen \cite{B:68}, Hayward \cite{H:06}, and Dymnikova \cite{D:92} metrics — illustrative examples that have inspired extensive study — but the literature contains many more \cite{LZ:11,BM:13,SV:19,RS:18}. Lacking a derivation from first principles, these geometries are limited to providing \textit{kinematical} insights. As a result, research has increasingly turned toward two central questions: how such regular solutions can be sourced or dynamically generated, and what observable features they may possess. 

A major challenge in this study has been realizing regular geometries as exact solutions of a fundamental theory. Nonlinear electrodynamics were initially conceived with the intent to cure the divergences associated with the electric field self-energy of charged particles present in the linear Maxwell theory \cite{B:33,BI:34a,BI:34b}. However, it has been shown to offer one possible avenue in enabling a wide class of regular metrics to be generated \cite{pt:69,bmss:79,ag:98,ag:99a,ag:99b,b:00,b:01,bh:02,d:04,bv:14,fw:16,b:17,tsa:18,b:23}. Nevertheless, this approach introduces significant problems: the Lagrangians often lack a Maxwell limit, requiring exotic matter; they can be acausal; and they demand \textit{fine-tuned} parameters, making regular solutions non-generic \cite{fw:16,ms:24}. Even in the simplest cases, singularity resolution is not a robust feature of the theory. While alternative constructions \cite{CM:21,LL:24} alleviate some concerns, they still depend on matter couplings, and the singular Schwarzschild solution remains when matter is absent.

A more promising direction has emerged within modified gravity theories, particularly quasitopological gravity. In dimensions $D\geqslant 5$, the inclusion of an infinite tower of higher-curvature corrections to the Einstein–Hilbert action has been shown to resolve singularities in a generic fashion \cite{BCH:25}. Quasitopological theories possess a number of desirable features: they provide a basis for effective field theory in vacuum, yield second-order field equations in spherical symmetry, and satisfy a Birkhoff theorem — ensuring the static solution is unique \cite{BCH:19, BCHLM:22, MM:23,BHM:25}. The connection between singularity resolution and the resummation of an infinite curvature expansion is particularly appealing, as such towers are a common prediction across various quantum gravity approaches. Crucially, regularity is achieved \textit{without} introducing additional matter fields. This framework arguably constitutes the first comprehensive approach to the regular black hole problem, enabling dynamical studies \cite{BCHM:25a, BCHM:25b, BCHMC:25} while guaranteeing that even in the absence of matter, the metric remains singularity-free.

It is important to stress, however, that such theories are non-trivial only in dimensions $D\geqslant 5$ and become trivial in $D=4$ \cite{BCH:19,BCHLM:22,BCH:25}. In four dimensions, they reduce to Einstein gravity and therefore cannot regularize the geometry. This represents a significant limitation, particularly since our universe is believed to be four-dimensional.\footnote{Derivations in four dimensions have been performed within the context of four-dimensional scalar-tensor theories obtained from a well-defined dimensional regularization of Lovelock invariants in Ref.~\cite{F:25}. However, in that work, the integration of the equations of motion was only feasible for the case of planar black holes. Another interesting derivation is given in Ref.~\cite{BCR:26}, where regular objects are derived within four-dimensional \textit{non-polynomial} quasitopological gravity. More generally, the relation between integrable two-dimensional
	Horndeski/dilaton theories and $D\geqslant 4$ gravitational theories with
	Birkhoff-type static vacuum solutions has recently been clarified in
	Ref.~\cite{B:26}. For related applications of four-dimensional non-polynomial
	quasitopological gravities to non-singular cosmology, see Ref.~\cite{BM:26}.} An alternative and well-developed approach, directly connected to quantum gravity -- specifically loop quantum gravity -- has been proposed in the literature. In this framework, one begins with a modified scalar constraint incorporating quantum gravity corrections, which in turn leads to unique static solutions in spherical symmetry. Furthermore, there exists an effective description of this quantum gravity framework within the context of mimetic gravity \cite{Chamseddine:2016uef,Langlois:2017hdf,BenAchour:2017ivq,Han:2020uhb,Han:2022rsx,GLSW:25,GL:25,BenAchour:2018khr}, providing a useful way to capture the modified dynamics at the covariant level.

This connection is important for the covariant interpretation of the construction. The dust, or equivalently a constrained scalar reference field, is introduced to deparameterize the system: it provides a relational clock, defines the preferred foliation, and allows the modified spatial Hamiltonian to be promoted to a four-dimensional covariant theory. It should not be interpreted as physical matter sourcing the regular vacuum geometry. On the vacuum branch considered here, the dust density that couples to the deparameterized system is set to zero; the remaining reference congruence acts only as a test-observer clock. In this sense, the covariant scalar sector carries no independent physical degree of freedom in the vacuum sector. The physical dynamics are generated by the deparameterized Hamiltonian rather than by an additional propagating scalar mode.

Since the covariant theory is reconstructed from this spatial Hamiltonian using the constrained clock field to define the foliation, the higher-curvature terms arise through projected curvature and extrinsic-curvature combinations rather than through unconstrained higher time derivatives. This is the mechanism by which the construction avoids introducing Ostrogradsky ghost instabilities in the sector under consideration. It also distinguishes the present model from standard mimetic-gravity constructions in which the scalar sector can represent an additional degree of freedom and may be associated with stability issues. Thus the scalar is retained explicitly as an auxiliary reference field: its derivatives define the normal vector and spatial projector needed to rewrite the spatial Hamiltonian in four-dimensional covariant form, including the resulting projected-curvature and extrinsic-curvature couplings, but it is not introduced as a new matter component sourcing the vacuum geometry.

Regularization in ultracompact objects is typically realized through a de Sitter (dS) core --- a feature shared by models such as Bardeen, Hayward, and Dymnikova. However, regularization can in principle also be achieved by introducing an anti--de Sitter (AdS) core. In either case, a regular geometry cannot satisfy all the classical energy conditions everywhere: at least one of the assumptions entering the singularity theorems must fail \cite{HE:73,Bambi:book:23}. The precise energy condition that is violated, and the region in which the violation occurs, are model-dependent.

For reference, the strong energy condition (SEC) in $D$-dimensional GR takes the form
\begin{align}
	\left(T_{\mu\nu}-\frac{1}{D-2}g_{\mu\nu}T\right)t^{\mu}t^{\nu}\geqslant 0,
\end{align}
where $T_{\mu\nu}$ is the energy-momentum tensor, $T$ is its trace, and the inequality must hold for every timelike vector $t^{\mu}$. A dS core is characterized by positive energy density and negative pressure, and typically violates the SEC near the center. By contrast, an exact AdS-like core has negative effective energy density and positive pressure; it therefore violates the weak energy condition at the center, while the SEC need not be violated there \cite{ALNV:25}. It has been shown that as a star is compressed toward the Buchdahl limit \cite{B:59}, its internal pressure grows and may become arbitrarily large and positive. This can trigger significant quantum effects, whose backreaction generates an effective energy–momentum tensor admitting negative energy densities near the center \cite{ABRG:22,ABRG:24,RT:23,R:23} \footnote{The AdS core scenario has been shown to arise within quasitopological gravity when coupled to Maxwell theory, as well as in nonlinear electrodynamics. In the latter case, a transition from a dS to an AdS core occurs when the electrostatic self-energy exceeds the
	Arnowitt-Deser-Misner (ADM) mass \cite{HKMS:25}. Effective energy conditions are analyzed in Appendix B of Ref.~\cite{HKMS:25}, revealing that the regions of violation differ between the two cases. Specifically, while the SEC is violated near the center for a dS core, in the AdS case the violation can extend further away from the center. Nevertheless, the characteristic feature common to both scenarios is the violation of energy conditions in some region.}. From this perspective, the AdS-core scenario is well motivated. Recently, it has been shown that there is a consistent way to modify existing regular metrics — originally constructed with dS cores — so that they instead admit AdS cores \cite{ALNV:25}. 

The outcome of this process constitutes the central aim of the present work. The remainder of this paper is organized as follows: In Sec.~\ref{sec:dS-adS}, we briefly review the method for obtaining regular solutions with an AdS core from those originally constructed with a dS core. While the AdS-core interpolation procedure is generic, we choose a specific Hayward-based representative for calculational simplicity and use it to illustrate the main results of this paper. In Sec.~\ref{sec:LTB}, we quickly review the process of deriving polymerized vacuum solutions from Lemaître–Tolman–Bondi (LTB) collapse in the non-marginally bound case and describe how to generate the aforementioned AdS-core Hayward model. In Sec.~\ref{sec:bounce}, we analyze bounce scenarios and study how the presence of a bounce is encoded, or missed, in finite-order effective truncations of the full model. In Secs.~\ref{sec:Hamiltonian} and \ref{sec:Lagrangian}, we derive the physical Hamiltonian for a specific model and subsequently obtain the four-dimensional covariant Lagrangian, respectively. Finally, in Sec.~\ref{sec:discussion} we summarize the most important results of our work and indicate potential future directions. Throughout this article, we restrict ourselves to four spacetime dimensions and we work in dimensionless units such that $\hbar=c=G=1$.\\

\section{From de Sitter to anti-de Sitter regularization}\label{sec:dS-adS}

\begin{figure*}[!htbp]
	\centering
	\subfloat[Metric functions]{
		\includegraphics[scale=0.4]{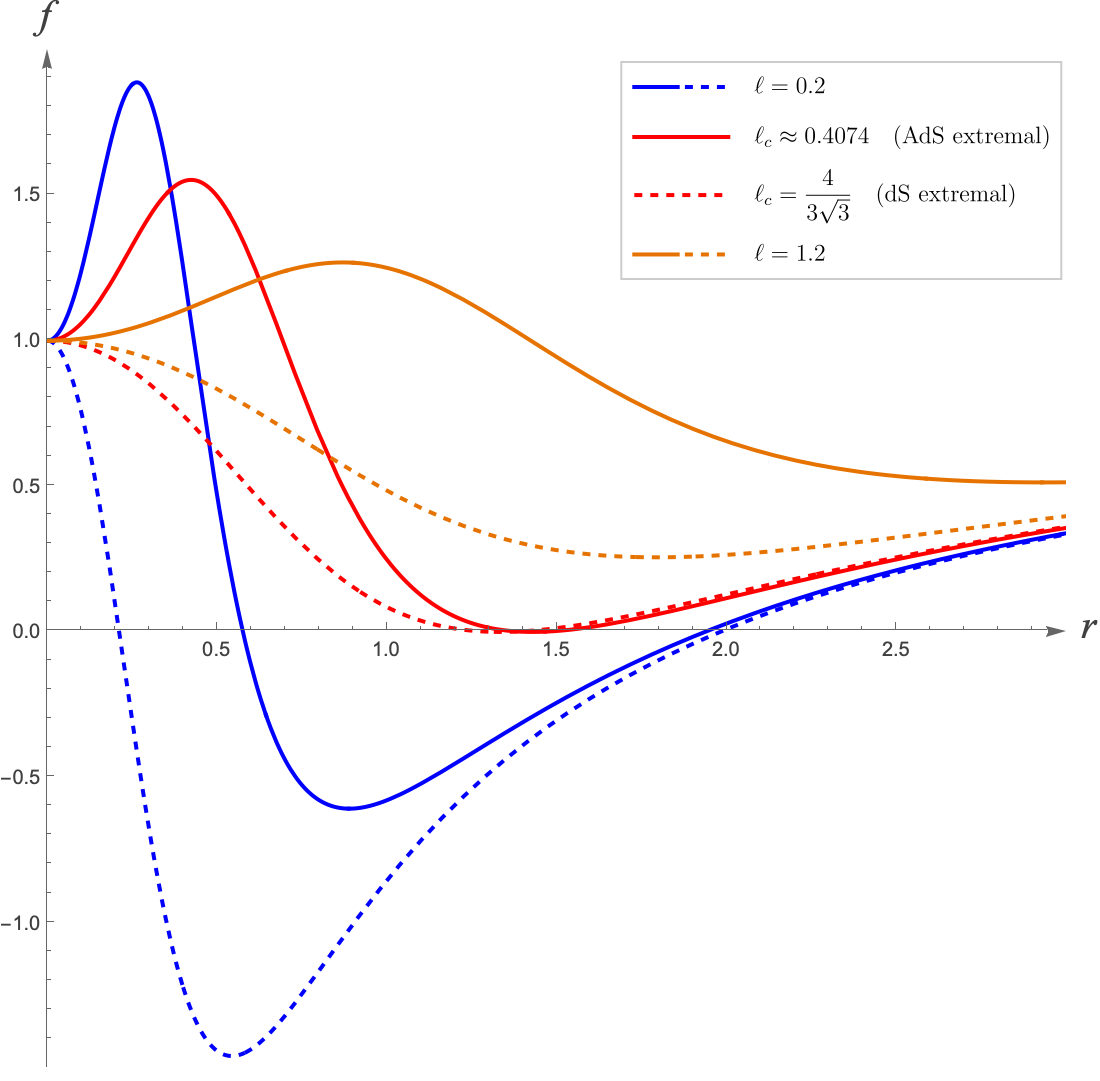}\label{fig:subfig:f}
	}
	\hspace{0.05\textwidth}
	\subfloat[Misner–Sharp mass]{
		\includegraphics[scale=0.4]{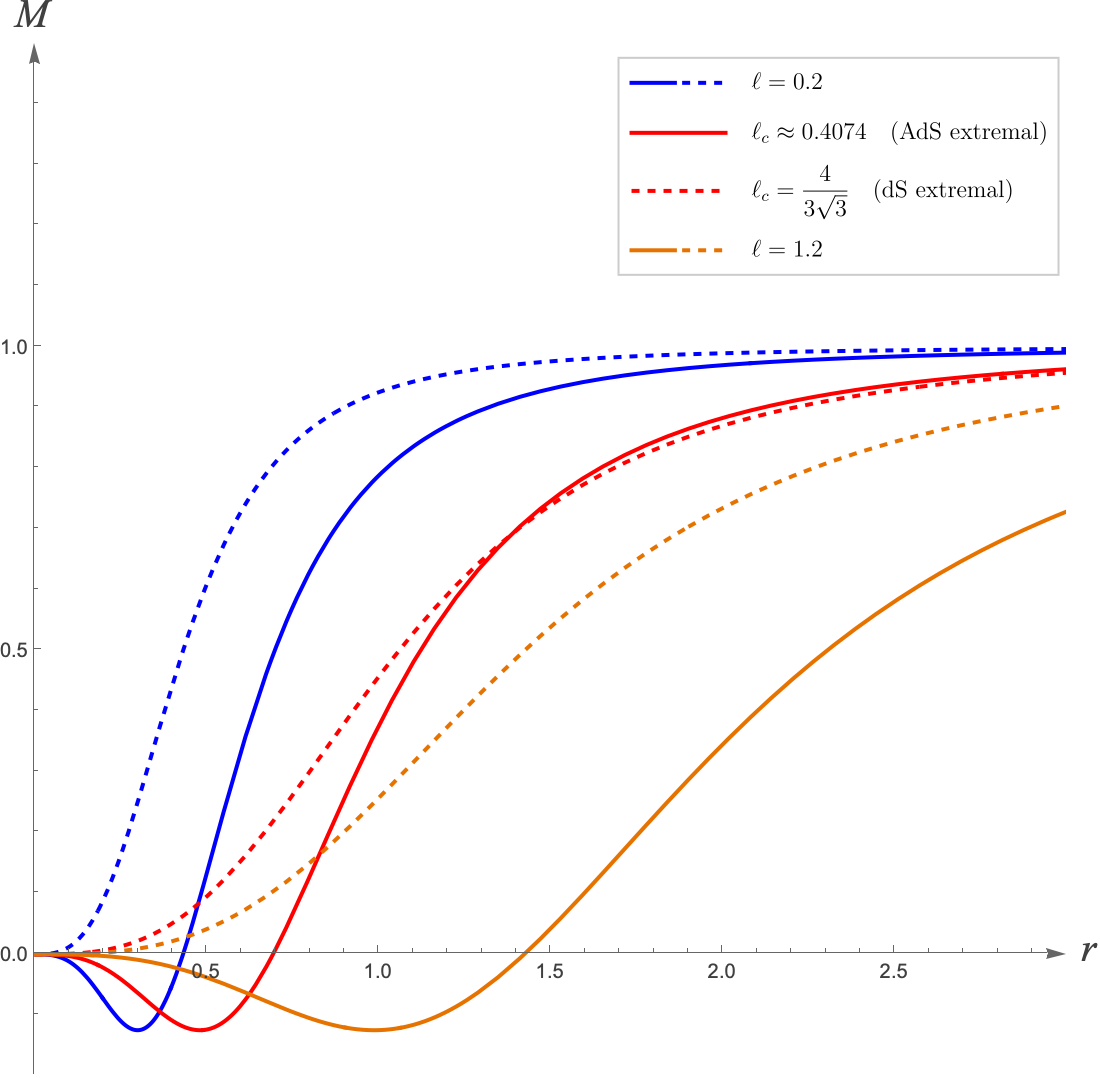}\label{fig:subfig:MS}
	}
	\caption{\textbf{Metric functions and MS mass for the Hayward model: dS and AdS cores.} (a) Metric functions $f(r)$ from Eqs.~\eqref{eq:fds} (dS core) and \eqref{eq:fads} (AdS core) plotted against the areal radius $r$ for different values of the minimal length scale $\ell$ with the ADM mass fixed to $m=1$. Dashed lines correspond to the dS-core metric, while solid lines represent the AdS-core metric with $\epsilon=1$, chosen to highlight the differences between the two cases. Blue curves correspond to black hole geometries featuring two horizons (where $f(r)=0$) for $\ell=0.2$. Orange curves represent horizonless configurations. Red curves depict extremal configurations, which occur at different critical minimal length scales due to the presence of $\epsilon$: the dashed red line shows an extremal dS-core black hole at $\ell_c=4/(3\sqrt{3})$, while the solid red line shows an extremal AdS-core black hole at $\ell_c\simeq 0.4074$. Both extremal cases exhibit a single degenerate horizon at distinct finite radii; (b) Corresponding MS mass $M(r)$ from Eqs.~\eqref{eq:MS-dS} and \eqref{eq:MS-AdS} for the same configurations. The same parameter values, line styles, and color coding apply as in panel (a). The plot illustrates the sign change near the center for the AdS-core case, occurring at a finite radius $r_{0}=(2 m \ell^2)^{1/3}$, and its absence in the dS-core case.}
	\label{fig:f}
\end{figure*}
The geometry of a general static spherically symmetric metric can be written in Schwarzschild coordinates as
\begin{align}
	ds^2 = -e^{2h(r)}f(r)dt^2_s + \frac{dr^2}{f(r)} + r^2 d\Omega^2,
\end{align}
where $r$ is the areal radius and the function $f(r)$ can be expressed in terms of the Misner–Sharp (MS) mass $M(r)$ as
\begin{align}
	f(r) = 1 - \frac{2M(r)}{r}.
\end{align}
In this paper, we focus primarily on the case where $h(r)$ vanishes, i.e., $h(r)=0$. 

Following Ref.~\cite{ALNV:25}, one can transform a regular metric with a dS core into a regular metric with an AdS core as follows. Starting from a dS-core metric, we have
\begin{align}
	f(r) = 1 - \frac{2M_{\mathrm{dS}}(r)}{r}.
\end{align}
Near the center, the dS core implies a positive MS mass in that region, whereas an AdS core requires a negative MS mass there. To enable this sign change at some finite radius, we introduce a modulating function $\chi(r)$ and write the modified metric function as
\begin{align}
	f(r) = 1 - \frac{2M_{\mathrm{dS}}(r)\chi(r)}{r}. \label{eq:fmod}
\end{align}
The function $\chi(r)$ must satisfy two asymptotic conditions: first, it should preserve the Schwarzschild behavior at infinity, and second, it must induce the sign change that produces an AdS core near the origin. These requirements translate to
\begin{align}
	\lim_{r\to\infty} \chi(r) = 1 \quad \text{and} \quad \lim_{r\to 0} \chi(r) = -1. \label{eq:chi-limits}
\end{align}
A family of functions can interpolate between these two limits. Here we adopt the specific choice introduced in Ref.~\cite{ALNV:25}, as it proves particularly convenient for our analysis. This function is given by
\begin{align}
	\chi(r) = \frac{r^n - \epsilon \kappa^{n-p} \ell^p}{r^n + \epsilon \kappa^{n-p} \ell^p}, \qquad n \geqslant 1, p \geqslant 1, \label{eq:chi}
\end{align}
where $\kappa$ is a positive constant with dimensions of length, and $\ell$ is the regularization parameter, often interpreted as the characteristic length scale at which quantum-gravity corrections to Einstein's equations become relevant \cite{F:16}. The particular rational form of $\chi(r)$ in Eq.~\eqref{eq:chi} is not unique, but rather a convenient representative of the broader class of interpolating functions satisfying Eq.~\eqref{eq:chi-limits}. Different choices of $\chi(r)$ would define different AdS-core completions with the same required near-core and asymptotic limits. We also emphasize that $\ell$ should not be identified a priori with the Planck length $\ell_p$ \cite{COS:22,CLMMOS:23}\footnote{Note that this length scale may vary. Studies incorporating backreaction effects in the black hole interior suggest an evolving minimal length $\ell$ \cite{BBCRG:21,BBCRG:22,CRFLV:23}. This motivates our allowance for variations in $\ell$ when discussing thermodynamics \cite{SS:24,S:24} or observational signatures \cite{ms:24}.}. In Eq.~\eqref{eq:chi}, the parameter $\epsilon$ is dimensionless and satisfies $\epsilon\geqslant 0$. The case $\epsilon=0$ recovers the original dS-core metric, while small positive values of $\epsilon$ correspond to AdS-core metrics that are close to their dS counterparts. Moreover, $\epsilon$ must remain non‑negative to preserve the regularity of the metric function.

\begin{figure*}[!htbp]
	\centering
	\subfloat[Energy density $\rho$]{
		\includegraphics[width=0.29\textwidth]{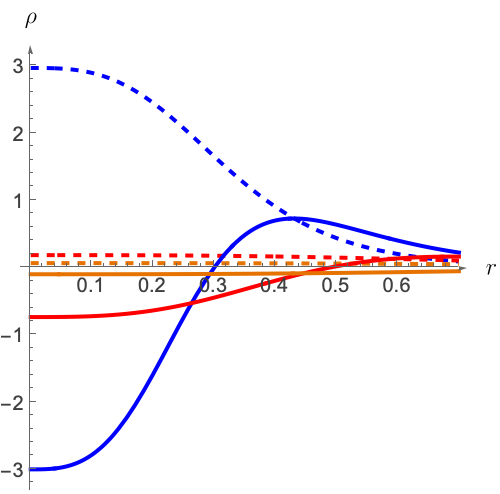}\label{fig:subfig:rho}
	}
	\hspace{0.02\textwidth}
	\subfloat[Radial pressure $p_r$]{
		\includegraphics[width=0.29\textwidth]{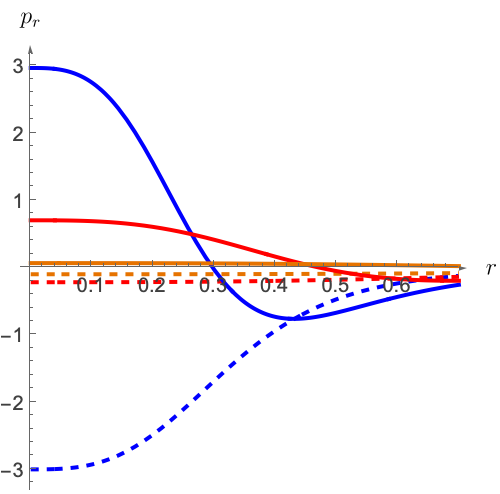}\label{fig:subfig:pr}
	}
	\hspace{0.02\textwidth}
	\subfloat[Tangential pressure $p_t$]{
		\includegraphics[width=0.29\textwidth]{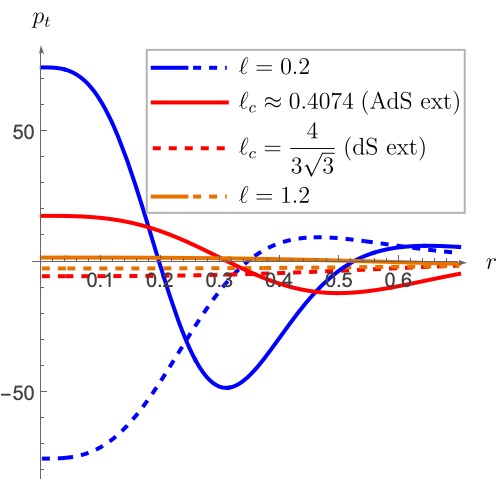}\label{fig:subfig:pth}
	}
	\caption{\textbf{Energy density and pressures.} In this figure we use the same models as in Fig.~\ref{fig:f}, with identical parameter values, line styles, and color coding. (a) The energy density $\rho$ as a function of the areal radius. The dashed lines (dS cores) are positive near the center, while the AdS cores exhibit the opposite behavior; (b) The radial pressure $p_r$, showing the reverse situation compared to the energy density: dS cores have negative pressure (tension), whereas AdS cores have positive radial pressure. (c) The tangential pressure $p_t$, which behaves similarly to the radial pressure for both scenarios.}
	\label{fig:rho}
\end{figure*}

To demonstrate the construction, we adopt the Hayward metric as our working example; this same model will be used throughout the paper. This model, which as we will show admits a dS core, is defined by the metric function
\begin{align}
	f_{\mathrm{dS}}(r)=1-\frac{2mr^2}{r^3+2m\ell^2}, \label{eq:fds}
\end{align}
where the MS mass is given by
\begin{align}
	M_{\mathrm{dS}}(r)=\frac{mr^3}{r^3+2m\ell^2},\label{eq:MS-dS}
\end{align}
and remains always positive. In the limit where the minimal length scale $\ell$ vanishes, the metric reduces to the singular Schwarzschild solution. Expanding the metric function \eqref{eq:fds} about the center, we find that
\begin{align}
	f_{\mathrm{dS}}(r)=1-\frac{r^2}{\ell^2}+\mathcal{O}(r^5),
\end{align}
which indicates the presence of a dS core. We now attempt to modify the metric to obtain an AdS core by employing the function $\chi(r)$ from Eq.~\eqref{eq:chi} with the choices $n=3$, $p=2$, and $\kappa=2m$\footnote{This choice is made for calculational convenience. In particular, it matches the scale \(2m\ell^2\) already appearing in the Hayward denominator, so that the resulting metric function keeps a simple dependence on the combination $2m/r^3$. This simplification is useful later when reconstructing and integrating the physical Hamiltonian $H_0$. The choice is not unique; other interpolating functions satisfying the same limiting conditions would define different AdS-core completions.}. With this modification, the new metric function obtained from Eq.~\eqref{eq:fmod} becomes
\begin{align}
	f_{\mathrm{AdS}}(r)=1-\frac{2mr^2}{r^3+2m\ell^2}\frac{r^3-\epsilon 2m\ell^2}{r^3+\epsilon 2m\ell^2}. \label{eq:fads}
\end{align}
Expanding Eq.~\eqref{eq:fads} about the center, we find that
\begin{align}
	f_{\mathrm{AdS}}(r)=1+\frac{r^2}{\ell^2}+\mathcal{O}(r^5),
\end{align}
which is clearly an AdS core. We also observe that the modified MS mass is given by
\begin{align}
	M_{\mathrm{AdS}}(r)=\frac{mr^3}{r^3+2m\ell^2}\frac{r^3-\epsilon 2m\ell^2}{r^3+\epsilon 2m\ell^2},\label{eq:MS-AdS}
\end{align}
which clearly admits a sign change at a finite radius $r_{0}=(\epsilon 2m\ell^2)^{1/3}$, effectively becoming negative for $r<r_{0}$. The comparison of the dS and AdS metric functions for various minimal length scales can be seen in Fig.~\ref{fig:subfig:f}. We can see the completely different behavior of the functions near the center. The solid lines which represent the AdS case lead to an initially increasing function of $r$, while the situation is completely different for the dS case represented by the dashed lines. This deviation is more pronounced due to our choice of $\epsilon=1$. If we had chosen $\epsilon\ll 1$, the AdS geometry would differ from the dS core case only in a region extremely close to the center. Finally, we should also mention once again that the MS mass changes sign for the AdS core case, which is in sharp contrast with the dS case. Solving $M_{\mathrm{AdS}}(r)=0$ yields $r_{0}=(2m\ell^2)^{1/3}$. This behavior can be clearly seen in Fig.~\ref{fig:subfig:MS}.

Before we proceed with the study of polymerized vacuum solutions, it is instructive to briefly examine the behavior of the corresponding energy density and pressures. For this analysis, it is convenient to assume that these metrics are solutions of the Einstein equations in GR sourced by some well-defined matter content described by an effective energy-momentum tensor $T_{\mu\nu}$. Namely, we define
\begin{align}
	\tensor{T}{^\mu_\nu}
	\equiv \frac{1}{8\pi}\tensor{G}{^\mu_\nu}.
\end{align}
In this spherically symmetric setting, the energy density and pressures as measured by a static observer are given by
\begin{align}
	\rho=-\tensor{T}{^t_t}, \quad p_{r}=\tensor{T}{^r_r}, \quad p_{t}=\tensor{T}{^\theta_\theta}=\tensor{T}{^\phi_\phi},
\end{align}
with the remaining components vanishing due to the absence of fluxes, as required by static spherical symmetry. These quantities should be understood as the effective GR stress tensor associated with the metric, and not necessarily as a fundamental matter source. In the modified-gravity interpretation adopted later, they simply provide a useful way of characterizing the geometric corrections responsible for the regular core.

The relevant plots can be found in Fig.~\ref{fig:rho}. Specifically, we observe in Fig.~\ref{fig:subfig:rho} that for regular solutions admitting a dS core, the energy density is positive, while in the AdS case -- which is our main interest -- the energy density becomes negative near the center and changes sign at a finite radius $r_{\star}$. As anticipated, the radial pressure exhibits the opposite behavior, since $p_r = -\rho$ as can be seen from Fig.~\ref{fig:subfig:pr}. This yields negative pressure (tension) for dS cores, a well-known feature in modern cosmology, and positive radial pressure for AdS cores. The tangential pressure $p_t$, shown in Fig.~\ref{fig:subfig:pth}, behaves similarly to the radial pressure and plays a crucial role in the violation of energy conditions in these spacetimes. For a detailed discussion, we refer the interested reader to Ref.~\cite{ALNV:25}.

Although the above construction is framed within GR, the effective energy-momentum tensor responsible for regularization can be reinterpreted as arising from the geometric sector of a modified gravity theory. This perspective is particularly useful when such corrections are encoded in additional terms in the Lagrangian. Prominent examples include quasitopological gravity and extended mimetic gravity models, the latter of which we will explore in the following sections. Before doing so, however, we first provide a brief review of the LTB collapse process and the generation of polymerized vacuum solutions.

\section{LTB collapse and polymerized vacuum solutions}\label{sec:LTB}

We begin with a brief review of the LTB formalism within GR to extract the features relevant for our analysis. These features will then be employed in the construction of modified theories intended as an effective description of quantum gravity. As most of the technical details are already available in the literature, we will not reproduce them exhaustively here. For a comprehensive exposition, we refer the reader to Refs.~\cite{LLB:07,GLRSW:25,GLSW:25}.

\subsection{Classical LTB model}

We start by considering a spherically symmetric spacetime coupled to pressureless dust. In terms of the Ashtekar-Barbero variables adapted to spherical symmetry, the phase space is spanned by the canonical pairs $(E^{x},K_{x})$ and $(E^{\phi},K_{\phi})$, where $E^{I}$ are the densitized triad components and $K_{I}$ are related to the extrinsic curvature\footnote{On a time slice $t=\mathrm{const}$, with an induced metric $h_{ab}$, we can introduce triads $e^{i}_{a}$ such that $h_{ab}=e^{i}_{a}e^{j}_{b}\delta_{ij}$. These triads introduce a local $\mathrm{SO}(3)$ gauge invariance, reflecting the local Lorentz invariance of the tetrad formalism. A triad version of the extrinsic curvature can also be defined via $k^{a}_{i}e^{i}_{b}=k_{ab}$. The inverse triads are then defined as $E^{a}_{i}=\frac{1}{2}\epsilon_{ijk}\epsilon^{abc}e^{j}_{b}e^{k}_{c}$ \cite{Rov:10,RV:15,T:10}.}. The general line element is
\begin{align}
	ds^2=-N^2dt^2+\frac{(E^{\phi})^2}{E^{x}}(dx+N^{x}dt)^2+E^{x}d\Omega^2,
\end{align}
where $N(t,x)$ and $N^{x}(t,x)$ are the lapse and shift functions respectively. Pressureless dust is a perfect fluid with negligible pressure and is therefore described by an energy-momentum tensor of the form
\begin{align}
	T_{\mu\nu}=\varrho u_{\mu}u_{\nu}. 
\end{align}
A series of convenient choices simplifies our analysis considerably. First, we align the spatial coordinate lines with the dust geodesics. This is reflected in the choice of the four-velocity, which takes the form $u^{\mu}=\delta^{\mu}_{t}$. The next step is to adopt the comoving gauge, i.e., $N=1$ and $N^{x}=0$, which leads to the metric
\begin{align}
	ds^2=-dt^2+\frac{(E^{\phi})^2}{E^{x}}dx^2+E^{x}d\Omega^2.\label{eq:metric}
\end{align} 

The coordinates $(t,x)$ can be interpreted as comoving with the dust, meaning that radially free-falling observers follow geodesics along $x=\mathrm{const}$, with the coordinate time coinciding with the proper time on these geodesics. We restrict this description to regions where the reference congruence remains regular, so that the dust time and shell label provide a good coordinate system. Possible shell crossings or caustics would indicate a breakdown of this reference-clock chart, rather than a physical dust instability, since the dust density is set to zero on the vacuum branch considered below. As will be shown later, constructing a 4D effective Lagrangian requires expressing it in terms of the eigenvalues of the extrinsic curvature $K_{ab}$ and the intrinsic 3-curvature ${}^{(3)}R_{ab}$. To this end, we define the following functions

\begin{align}
	\mathcal{K}_{1}(t,x)\equiv \tensor{K}{^\theta_\theta}=\frac{\dot{E}^{x}}{2E^{x}}, \quad \mathcal{K}_{2}(t,x)\equiv \tensor{K}{^x_x}+\tensor{K}{^\theta_\theta}=\frac{\dot{E}^{\phi}}{E^{\phi}}, \label{eq:Ks}
\end{align}
where the former represents the angular expansion rate\footnote{In isotropic cosmologies (Friedmann-Lemaître-Robertson-Walker) these two expressions are identical; however, this is not the case when a central object is present.}. Similarly, for the 3-Ricci tensor we have
\begin{align}
	\mathcal{R}_1(t,x)\equiv -\tensor{R}{^{\theta\phi}_{\theta\phi}}, \quad \mathcal{R}_{2}(t,x)\equiv \tensor{R}{^{x\theta}_{x\theta}}.
\end{align}
The former is the radial curvature, while the latter is the angular curvature, which measures the curvature of the 2-spheres. These geometric quantities will serve as the foundation for the classical dynamics formulation, as well as for any modifications we wish to explore.

The LTB metric is given by
\begin{align}
	ds^2=-dt^2+\frac{(R'(t,x))^2}{1+\mathcal{E}(x)}dx^2+R(t,x)^2d\Omega^2.\label{eq:LTB:metric}
\end{align}
Demanding that the metric above coincide with that of Eq.~\eqref{eq:metric} necessarily imposes a condition on the functions $E^{\phi}$ and $E^{x}$. We designate this as the LTB condition, and it is readily verified to be
\begin{align}
	(E^{x})'=2\sqrt{1+\mathcal{E}(x)}E^{\phi}.
\end{align}
From the equations of motion we can show that the function $\mathcal{E}(x)$ is a time-independent integration function that has both geometrical interpretation but also an inherent and intuitive physical meaning. By direct calculation of  $\mathcal{R}_{1}(t,x)$ and $\mathcal{R}_{2}(t,x)$ function representing the radial and angular curvature we can show that 
\begin{align}
	\mathcal{R}_{1}(t,x)=\frac{\mathcal{E}}{E^{x}}, \quad 	\mathcal{R}_{2}(t,x)=-\frac{\mathcal{E'}}{(E^{x})'}.
\end{align} 

The quantity $\mathcal{E}(x)$ has a clear physical interpretation: it represents the specific binding energy of the dust shell labeled by $x$. Depending on its value, three distinct cases can be identified:
\vspace*{0.2cm}
\begin{itemize}[noitemsep,nolistsep,leftmargin=12pt]
	\item Marginally bound case: $\mathcal{E}(x)=0$.
	This corresponds to a marginally bound shell. Intuitively, the shell has zero total energy and vanishing kinetic energy at infinity.
	\vspace*{0.1cm}
	\item Unbound case: $\mathcal{E}(x)>0$.
	The shell possesses positive total energy.
	\vspace*{0.1cm}
	\item Bound case: $-1<\mathcal{E}(x)<0$.
	The shell has negative total energy.
	\vspace*{0.2cm}
\end{itemize}

LTB solutions are characterized by two independent free functions. The first is $\mathcal{E}(x)$, introduced earlier, and the second is the MS mass $M(x)$, which represents the total energy enclosed within the dust shell labeled by the radial coordinate $x$. An important property of these functions is that they are time-independent. In particular, $\partial_t \mathcal{E} = 0$, a result that follows directly from Einstein’s equations for pressureless dust in comoving coordinates and is not an additional assumption. In fact, $\mathcal{E}(x)$ is one of two fundamental quantities conserved along each shell. The second is the Misner–Sharp mass $M(x)$, which likewise satisfies $\partial_t M = 0$ for dust. These conservation laws arise naturally in the comoving description, where there is no flux of matter or energy across the worldlines of the shells. As a result, each shell evolves independently, behaving as an isolated system. This independence is reflected in the dynamical evolution equation given by
\begin{align}
	\dot{R}^2=\frac{2M(x)}{R(t,x)}+\mathcal{E}(x).
\end{align}
This relation holds independently for each value of $x$, reflecting the hallmark feature of decoupling in LTB dynamics. Before moving on to theories beyond GR, it is useful to formulate this decoupling property within the Hamiltonian framework. To this end, for each shell labeled by $x$, we introduce new canonical variables $b$ and $v$, defined as
\begin{align}
	b(x)=\frac{K_{\phi}}{\sqrt{E^{x}}}, \quad v(x)=(E^{x})^{3/2}=R^3,
\end{align}
that satisfy the Poisson bracket
\begin{align}
	\left\{ b(x),v(y)\right\}=\frac{3}{2}\delta(x-y).
\end{align}
Therefore, the dynamics of each independent shell are generated by a Hamiltonian $H(b,v; \mathcal{E})$ associated with that shell. In this formulation, there is no coupling between different shells; each evolves independently, reflecting the decoupling property discussed earlier. Consequently, according to Hamilton’s equations, the time evolution of the canonical variables is given by
\begin{align}
	\dot{v}=\left\{v,H\right\}=\frac{3}{2}\partial_bH, \quad 	\dot{b}=\left\{b,H\right\}=-\frac{3}{2}\partial_{v}H. \label{eq:Poisson}
\end{align}
The consistency between these variables and the inverse triad components, together with the extrinsic curvature, is ensured once the LTB condition and its first derivative with respect to $x$ are imposed. Taking all these relations into account, one can show that the Hamiltonian can be written in the following form
\begin{align}
	H(b,v; \mathcal{E})=vH_{0}(b,\mathcal{R}_1).	\label{eq:H-factor}
\end{align}
The full Hamiltonian/scalar constraint $\mathcal{C}$ can be calculated from the shell Hamiltonian by
\begin{align}
	\mathcal{C}=\frac{\partial_{x}H(b, v; \mathcal{E})}{\sqrt{1+\mathcal{E}(x)}},
\end{align} 
where the LTB constraint has been used. 

An appropriate generalization can be carried out in the case of modified theories, under the minimal assumption that compatibility with LTB-type solutions is preserved. By this, we mean that there exists a subset of solutions of LTB type for which a decoupling occurs, allowing each shell, labeled by $x$, to be described independently. The overall procedure remains essentially the same; however, the modified dynamics are now governed by a new scalar constraint, which must reduce to the GR limit in the appropriate parameter regime.

In the dust model, a vacuum region is characterized by the vanishing of the dust energy density. More precisely, by vacuum we mean that the dust density that couples to the deparameterized system is set to zero, so that the dust does not source the metric. From the Hamiltonian perspective, this corresponds to a region in which the shell Hamiltonian, and hence the Misner--Sharp mass $M(x)$, is independent of the shell label $x$, so that $M(x)=\mathrm{const.}$. The energy function $\mathcal{E}(x)$ is not fixed by the vacuum condition itself; it is conserved along each shell but may vary from shell to shell, encoding the choice of LTB reference congruence/slicing. In what follows, and in particular in Sec.~\ref{sec:bounce}, we exploit the decoupling of the shells
and study representative vacuum shell sectors with fixed values of $\mathcal{E}$, corresponding to the marginally bound, unbound, and bound cases.

We should stress that, in the context of a modified theory characterized by a modified scalar constraint, the term ``vacuum solution'' refers to a modified vacuum configuration. In our framework, this is commonly referred to as a polymerized vacuum, reflecting its relation to loop quantum gravity. The reference dust field is nevertheless retained in the covariant action only as the relational clock and foliation variable needed to write the deparameterized Hamiltonian in four-dimensional form; on the vacuum branch it carries no independent physical degree of freedom and should not be interpreted as an additional matter source.

An important issue to clarify is how this construction yields unique static solutions. The procedure proceeds as follows: we start from the LTB metric in Eq.~\eqref{eq:LTB:metric} and perform a coordinate transformation to Painlevé–Gullstrand coordinates,
\begin{align}
	dr=\dot{R}dt+R'dx.
\end{align}
This leads to
\begin{align}
	ds^2=-dt^2+\frac{1}{1+\mathcal{E}}(dr-\dot{R}dt)^2+r^2d\Omega^2.
\end{align}
In these coordinates, we can immediately identify the shift vector as
\begin{align}
	N^{r}(t,r)=-\dot{R}, \label{eq:Nr-dot}
\end{align}
and we should note that the conserved quantities $\mathcal{E}(x)$ and $M(x)$ are now expressed as functions of $(t,r)$. 

Our next step is to transform this metric to Schwarzschild coordinates $(t_s,r)$ in order to determine whether vacuum solutions exist. To achieve this, we must eliminate the cross terms, which requires imposing the condition $g_{t_s r}=0$ after the coordinate transformation. The new time coordinate is taken to be a function $t_s=t_s(t,r)$. Combining these requirements leads to the condition
\begin{align}
	\partial_{r}t_s=\frac{N^{r}}{1+\mathcal{E}-(N^r)^2}\partial_{t}t_s.
\end{align}
This coordinate transformation leads to the metric
\begin{align}
	ds^2=-\frac{1+\mathcal{E}-(N^{r})^2}{(1+\mathcal{E})(\partial_tt_s)^2}dt^2_s+\frac{1}{1+\mathcal{E}-(N^{r})^2}dr^2+r^2d\Omega^2.
\end{align}
We immediately observe that, at this stage, the metric is not uniquely determined, since it still contains the arbitrary LTB energy function $\mathcal{E}$. A nonconstant $\mathcal{E}(x)$ should be understood as part of the choice of reference congruence/slicing, rather than as an obstruction to the existence of a timelike Killing vector in the underlying vacuum spacetime.
For the static reconstruction pursued here, we therefore choose a shell-independent energy representative, which removes this slicing ambiguity and allows the metric coefficients to depend only on $(M,r)$. To achieve this, we introduce the following ansatz
\begin{align}
	(N^{r})^2=\mathcal{E}+1-f(M,r).
\end{align}
Using this and imposing the condition $(1+\mathcal{E})(\partial_{t}t_s)^2=1$ then leads to a metric of the form
\begin{align}
	ds^2=-f(M,r)dt^2_s+\frac{dr^2}{f(M,r)}+r^2d\Omega^2.
\end{align}
This metric is static, admitting the Killing vector $\partial_{t_s}$, and is unique for a given $M=m$.

We now proceed by combining Eq.~\eqref{eq:Nr-dot} with Eq.~\eqref{eq:Poisson} for $\dot{v}$, from which we obtain
\begin{align}
	N^{r}=-\frac{1}{2r^2}\partial_{b}H,
\end{align}
which leads to the partial differential equation
\begin{align}
	(\partial_{b}H)^2=4r^4\left(1+\mathcal{E}-f(M,r)\right).
\end{align}
Using now the factorization property of the Hamiltonian from Eq.~\eqref{eq:H-factor}, together with the relation $\mathcal{E}=r^2\mathcal{R}1$, we obtain
\begin{align}
	\partial_{b}H_0=2\left(\mathcal{R}_1+\frac{1-f(M,r)}{r^2}\right)^{1/2}.
\end{align}
The left-hand side of this equation depends on $b$ and $\mathcal{R}_1$, and therefore the right-hand side must depend on the same variables. Furthermore, when considering vacuum shells, the Hamiltonian is related to the mass through the relation $2M=r^3H_{0}$. Combining these observations implies that the quantity $(1-f)/r^2$ must take a generic form given by
\begin{align}
	\frac{1-f(M,r)}{r^2}=\tilde{f}(H_0(b,\mathcal{R}_1)),
\end{align}
which can ultimately be recast as \footnote{We note that the formula used to determine the metric function $f$ bears a remarkable resemblance to the structure that arises in quasitopological gravity. In that context, the equations of motion reduce to a relation of the form $h(\psi)=S(r)$, where $\psi=(1-f)/r^2$ and $S(r)=2m/r^{D-1}$ in the absence of matter fields and a cosmological constant \cite{BCH:25,HKMS:25}. This leads to a metric function of the form $f(r)=1-r^2h^{-1}(2m/r^{D-1})$. The structure is therefore formally identical to the one obtained here, written in the form of Eq.~\eqref{eq:f-LTB}. The main distinction is that, in our case, the construction is valid in four dimensions, unlike polynomial quasitopological gravity. Moreover, the sign of the limit $\displaystyle \lim_{r \to 0} \tilde{f}(2m/r^3)$ determines the nature of the core: a positive value corresponds to a dS core, whereas a negative value signals an AdS core.}
\begin{align}
	f(r)=1-r^2\tilde{f}\left(\frac{2m}{r^3}\right).\label{eq:f-LTB}
\end{align}
This equation constitutes the central relation used to construct regular solutions with an AdS core, with $\tilde{f}$ capturing the modifications of the underlying theory.
	Let us make explicit the conditions on \(\tilde f\) that are needed for asymptotic flatness and regularity. A general curvature-regularity criterion for polymerized static metrics of the form \(f(r)=1-r^2\tilde f(u(r))\), including asymptotically AdS backgrounds, was given in Ref.~\cite{BS:26}. In the present asymptotically flat case, the relevant variable is
\begin{align}
X=\frac{2m}{r^3}.
\end{align}
First, Schwarzschild asymptotics require
\begin{align}
\tilde f(X)=X+\mathcal O(X^{1+\delta}),
\quad X\to0,
\quad \delta>0 .
\end{align}
Equivalently, \(\tilde f(0)=0\) and \(\tilde f'(0)=1\), so that
\begin{align}
f(r)=1-\frac{2m}{r}+\mathcal O(r^{-1-3\delta}),
\quad r\to\infty .
\end{align}

The regularity conditions have a direct geometrical origin. If we write
\begin{align}
F(r)=\tilde f\!\left(\frac{2m}{r^3}\right),
\qquad
f(r)=1-r^2F(r),
\end{align}
then the independent orthonormal frame curvature scales are controlled by
\begin{align}
F(r),\qquad rF'(r),\quad r^2F''(r).
\end{align}
Thus all scalar polynomial curvature invariants remain finite at the center if and only if these quantities remain bounded as \(r\to0^+\). Since \(r\to0^+\) corresponds to \(X\to\infty\), and
\begin{align}
rF'(r)=-3X\tilde f'(X),
\end{align}
\begin{align}
r^2F''(r)=9X^2\tilde f''(X)+12X\tilde f'(X),
\end{align}
the regularity criterion is equivalently the boundedness of
\begin{align}
\tilde f(X),\qquad X\tilde f'(X),\qquad X^2\tilde f''(X)
\end{align}
as \(X\to\infty\). Hence regularity is not obtained for an arbitrary deformation; it follows for smooth polymerization functions that saturate sufficiently mildly at large \(X\).

The limiting core geometry is obtained by imposing the stronger saturation conditions
\begin{align}
\tilde f(X)\to F_0,
\quad
X\tilde f'(X)\to0,
\quad
X^2\tilde f''(X)\to0,
\end{align}
as $ X\to\infty $.
Then
\begin{align}
f(r)=1-F_0r^2+\mathcal{O}(r^{2+a}), \quad a>0,
\end{align}
and the center is locally maximally symmetric at leading order, with
\begin{align}
R\to12F_0,
\quad
R_{\mu\nu\rho\sigma}R^{\mu\nu\rho\sigma}\to24F_0^2 .
\end{align}
Therefore \(F_0>0\) corresponds to a de Sitter core, \(F_0<0\) to an anti-de Sitter core, and \(F_0=0\) to a Minkowski core. In particular, an AdS core with radius \(\ell_{\rm AdS}\) corresponds to
\begin{align}
F_0=-\frac{1}{\ell_{\rm AdS}^2},
\quad
f(r)=1+\frac{r^2}{\ell_{\rm AdS}^2}+\mathcal{O}(r^{2+a}),
\end{align}
with $a>0$. For positive mass and Schwarzschild normalization, \(\tilde f(X)>0\) near \(X=0\). Therefore an AdS core requires \(\tilde f\) to become negative on the physical branch, which is the origin of the sign change of the local Misner--Sharp mass in the AdS-core solutions.
\color{black}
\section{To bounce or not to bounce?}\label{sec:bounce}

The Hayward solution with an AdS core can be obtained as a vacuum solution using the method described in Sec.~\ref{sec:LTB}. In particular, starting from the metric function given in Eq.~\eqref{eq:fads}, we determine that the function $\tilde{f}$, which encodes the modifications introduced by the underlying modified theory, is given by
\begin{align}
	\tilde{f}(X)=\frac{X}{1+\ell^2X}\frac{1-\epsilon \ell^2X}{1+\epsilon \ell^2 X},
\end{align}
where $X=2m/r^3$. We express the result in this form to highlight that the modifications are implemented solely as a function of $X$. As expected, in the limit of vanishing minimal length scale $\ell$, the expression reduces to $\tilde{f}=X$, corresponding to the Schwarzschild geometry. Although the reconstructed vacuum geometry is static, the LTB/deparameterized description represents it through a family of radial shell trajectories, or equivalently through a collapse slicing associated with the same vacuum spacetime. The bounce discussed below is therefore a turning point of these shell trajectories, not a time-dependent bounce of the static metric itself. In the vacuum branch considered here the dust density is set to zero, so the shells should be understood as reference-clock trajectories, or test observers, probing the polymerized vacuum solution rather than as physical matter sources. If the dust density is not set to zero, the same LTB shell dynamics instead describes physical dust matter whose collapse sources the geometry; it is also the relevant structure in cosmological slicings and dust-collapse models. With this interpretation, the analysis of the shell bounce is directly connected to our requirement of staticity, namely,
\begin{align}
	\dot{R}^2=\mathcal{E}+r^2\tilde{f}\left(\frac{2m}{r^3}\right). \label{eq:dR}
\end{align}
In order to avoid a bounce, we require that the expression on the right-hand side of the above equation remain non-negative for all values of the areal radius $r$. A more intuitive way to understand this condition is through the MS mass\footnote{Strictly speaking, a zero of the right-hand side of Eq.~\eqref{eq:dR} is a single-shell turning point. Defining \(F(r)=\mathcal{E}+r^2\tilde f(2m/r^3)=\mathcal{E}+2M(r)/r\), a collapsing shell undergoes an inner bounce when \(F(r_b)=0\) and \(F'(r_b)>0\), so that \(\ddot R>0\). In the vacuum branch considered here, however, the LTB/reference congruence describes a family of test observers used to define a relational or comoving chart; it does not represent physical dust matter sourcing the geometry. The bounce therefore refers to the turning point of an individual test-observer trajectory. This should be distinguished from shell crossing, which is a property of the reference congruence and occurs when the shell-label map ceases to be invertible, \(R'(t,x)=0\) at \(R>0\). Thus, a bounce does not by itself imply shell crossing. Avoiding shell crossing requires choosing the LTB/reference congruence so that \(R'\) remains nonzero in the region covered by the comoving chart. If this condition fails, the dust-time chart breaks down, but this does not describe the crossing of physical matter shells or a physical dust instability on the vacuum branch, where the dust density is set to zero.}. In particular, one can readily verify that Eq.~\eqref{eq:dR} can be rewritten as

\begin{align}
	\dot{R}^2=\mathcal{E}+\frac{2M(r)}{r}.
\end{align}

This means that interesting behavior might be more possible (although not necessarily as the sign of $\mathcal{E}$ plays  a role ) if there is a sign change for the MS mass. This is precisely the case in the Hayward model with AdS core (See Fig.~\ref{fig:subfig:MS}). Therefore, it will be interesting to check and compare the behavior of the bounce for dS and AdS core solutions. In the following we want to maximize the distinction between dS and AdS core cases and thus we pick the values $\epsilon=0$ and $\epsilon=1$ respectively in Eq.~\eqref{eq:MS-AdS}. 
\begin{figure*}[!htbp]
	\centering
	\includegraphics[width=\textwidth]{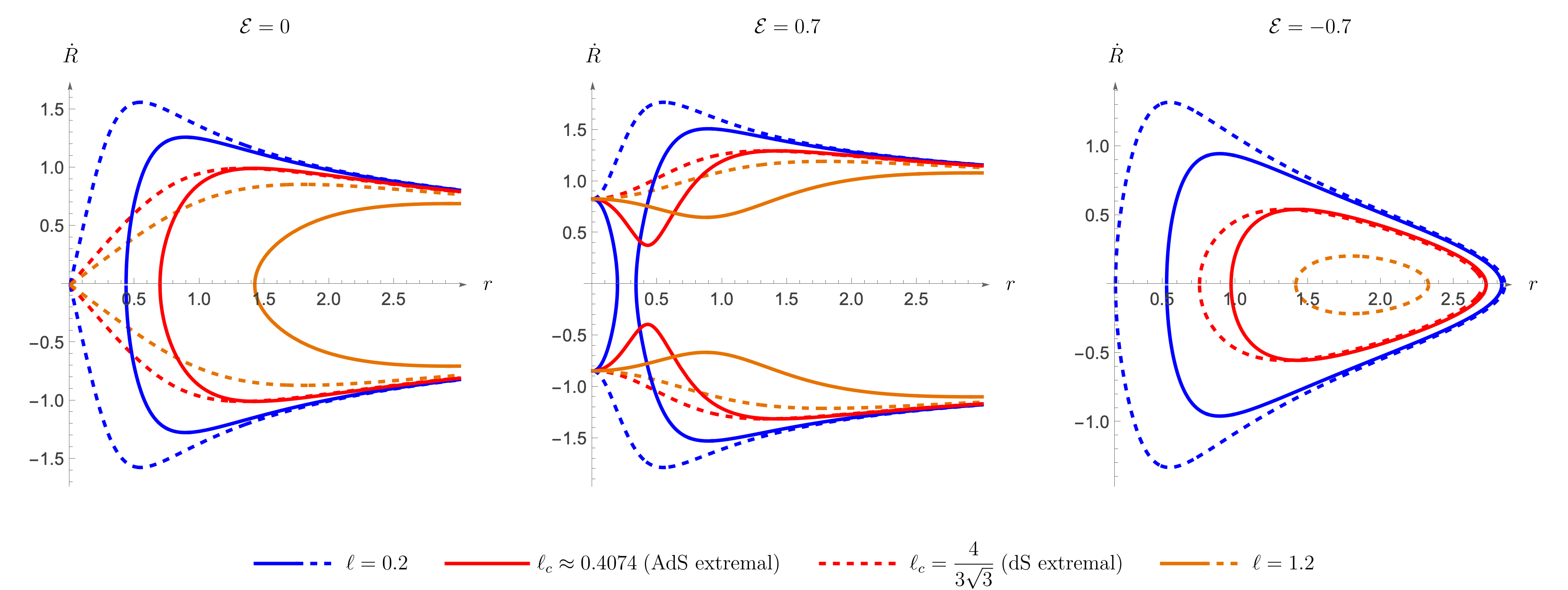} 
	\caption{\textbf{Bounces.} The real branches of $\dot{R}$ are plotted as functions of the areal radius $r$ for three representative cases: the marginally bound case $\mathcal{E}=0$ (left), the unbounded case $\mathcal{E}>0$ with $\mathcal{E}=0.7$ (middle), and the bounded case $\mathcal{E}<0$ with $\mathcal{E}=-0.7$ (right). The curves correspond to the same models as in Fig.~\ref{fig:subfig:f}, with identical parameter choices, line styles, and color coding. We note that in the right panel the solid orange curve is absent, since for this value of the minimal length scale the function $\dot{R}$ is not defined.}
	\label{fig:dotR}
\end{figure*}

We begin with the simplest and most commonly studied scenario: the marginally bound case, characterized by $\mathcal{E}=0$. In this case, the sign of the MS mass is the only factor determining the existence of a bounce. For the dS core case, it is clear that since the MS mass does not change sign, there will be no bounce with $\dot{R}$ vanishing exactly at the center $r=0$ in agreement with the analysis of Ref.~\cite{GLSW:25}. The situation is drastically different for the AdS case. As we saw in Sec.~\ref{sec:dS-adS}, the MS mass changes sign, indicating a region of negative energy density near the center. This sign change occurs at $r_0 = (2m\ell^2)^{1/3}$ (recall that we have set $\epsilon = 1$; otherwise, this radius depends on $\epsilon$. In the case $\epsilon = 0$, corresponding to a dS core, $r_0 = 0$ and no sign change occurs). Interestingly, as shown in Fig.~\ref{fig:dotR}, the bounce radius lies deeper inside the core for black hole configurations but occurs further away for horizonless configurations. This can be understood intuitively from the spatial distribution of negative energy density: it is more centrally concentrated in black holes, while in horizonless configurations it extends over a larger region. This is clearly visible in Figs.~\ref{fig:subfig:MS} and \ref{fig:rho}.

\begin{table*}[t]
	\centering
	\renewcommand{\arraystretch}{1.35}
	\setlength{\tabcolsep}{10pt}
	\begin{tabular}{|c||l||l|}
		\hline
		\rule{0pt}{2.6ex} $\mathcal{E}$ 
		& \multicolumn{1}{c||}{dS Core} 
		& \multicolumn{1}{c||}{AdS Core} \rule{0pt}{2.6ex}\\
		\hline\hline
		\rule{0pt}{2.6ex} $\mathcal{E}=0$ 
		& $\xmark$
		& $\cmark$ \rule{0pt}{2.6ex}\\
		\hline
		\rule{0pt}{2.6ex} $\mathcal{E}>0$ 
		& $\xmark$
		& $\cmark$ \quad ($\ell<\ell_0$ or $\mathcal{E}<\mathcal{E}_0$) \rule{0pt}{2.6ex}\\
		\hline
		\rule{0pt}{2.6ex} $-1<\mathcal{E}<0$ 
		& $\cmark$ \quad ($\ell<\bar{\ell}_0$ and $\mathcal{E}> \bar{\mathcal{E}}_{0}$)
		& $\cmark$ \quad ($\ell<\ell_0$ or $\mathcal{E}>\mathcal{E}_0$) \rule{0pt}{2.6ex}\\
		\hline
	\end{tabular}
	\caption{\textbf{Bounces.}  Existence of a turning point for different values of the shell energy $\mathcal{E}$ in the dS and AdS core cases. A turning point is indicated by $(\cmark)$, while its absence is indicated by $(\xmark)$. In the AdS case, the occurrence of a bounce in both the unbounded and bounded regimes is determined by the critical values $\ell_0$ and $\mathcal{E}_0$. In the dS case, a similar threshold behavior appears in the bounded regime, with the corresponding critical values denoted by $\bar{\ell}_0$ and $\bar{\mathcal{E}}_0$.}
	\label{tab:bounce}
\end{table*}

We now turn to the unbounded case, $\mathcal{E}>0$. In contrast to the bounded configurations, the analysis is more involved, since the dynamics depends not only on the sign of the MS mass but also explicitly on the magnitude of $\mathcal{E}$. The qualitative behavior of the shell is therefore highly sensitive to the choice of parameters. An important simplification occurs for the Hayward geometry with a dS core. In this case, the right-hand side of Eq.~\eqref{eq:dR} remains strictly positive for all $r$, owing to the positivity of both the MS mass and $\mathcal{E}$. Consequently, the shell does not experience a qualitative change in its motion. The situation is markedly different for the AdS core, where the dynamics depends nontrivially on the parameter space. We analyze this case in two complementary -- and physically equivalent -- ways.

First, we fix the shell energy to $\mathcal{E}=1>0$ and vary the minimal length scale $\ell$. We find the existence of a critical value $\ell_{0}<\ell_{c}$, where $\ell_{c}$ denotes the extremal black hole length scale, which separates bouncing from non-bouncing solutions. Specifically, for $\ell<\ell_{0}$ a bounce occurs, whereas for $\ell>\ell_{0}$ the shell collapses without reversal. Physically, for fixed $\mathcal{E}$ only sufficiently small values of $\ell$ allow for a bounce. This can be understood from Figs.~\ref{fig:subfig:MS} and \ref{fig:subfig:rho}, which show that smaller $\ell$ enhances the region of negative effective energy density. The resulting stronger negative energy contribution provides the necessary repulsive effect to reverse the motion of the shell. For larger $\ell$, this repulsion is insufficient and no bounce takes place.

Alternatively, one may fix $\ell$ and vary the shell energy $\mathcal{E}$. In this parametrization, there exists a critical value $\mathcal{E}_{0}$ separating the two dynamical regimes. In the bouncing regime, the equation $\mathcal{E}+2M(r)/r=0$ can have two positive roots, $r_{-}<r_{+}$, with a forbidden interval between them where $\dot{R}^{2}<0$. A shell collapsing from large $r$ encounters the outer root $r_{+}$ first, so this is the relevant bounce radius for the exterior trajectory; at the critical value the two roots merge. For $\mathcal{E}<\mathcal{E}_{0}$, corresponding to relatively low-energy shells, a bounce occurs. In contrast, for $\mathcal{E}>\mathcal{E}_{0}$ the shell energy is too large for the negative energy effects to counterbalance the collapse, and no bounce is possible. These two perspectives describe the same underlying physics and are illustrated in Fig.~\ref{fig:dotR}.

Lastly, we consider the bounded case characterized by $\mathcal{E}<0$. In this regime, both the dS and AdS cores exhibit qualitatively similar behavior. For an appropriate range of parameters, the collapsing shell reaches an inner turning point at a finite radius and subsequently expands until it encounters an outer turning point, since its energy is insufficient to escape to infinity. The outer turning point corresponds to the recollapse of a closed universe configuration. The main quantitative difference between the dS and AdS cores lies in the location of these turning points. In general, the inner bounce in the dS case occurs closer to the center than in the AdS case. This can be understood from the stronger effective repulsion induced by the AdS core, associated with the negative contribution to the mass function near the origin, which shifts the balance between attraction and repulsion outward and therefore moves the inner turning point to larger radii. Conversely, in the AdS case the outer turning point typically occurs at a smaller radius compared to the dS case, reflecting the modified global structure of the effective potential. More generally, if we fix the minimal length scale $\ell$, there exists a critical negative value $\mathcal{E}_0$ at which the inner and outer turning points coincide. For $\mathcal{E}<\mathcal{E}_0$, no turning points exist and the bounded motion disappears. Similarly, for fixed $\mathcal{E}<0$, there is a critical length scale $\ell_0$ at which the two turning points merge. For $\ell<\ell_0$ two bounces occur, whereas for $\ell>\ell_0$ no bounce is present. This behavior is depicted in Fig.~\ref{fig:dotR} and all the cases are summarized in Table~\ref{tab:bounce}.

Interestingly, the behavior of the Hayward model with an AdS core is strikingly similar to that of loop quantum gravity–inspired geometries, such as the model described in Ref.~\cite{GL:25}. In particular, the qualitative structure of the metric function and the resulting dynamical features exhibit remarkable agreement. This correspondence suggests that AdS-like cores may capture essential features of certain quantum-gravity regularization mechanisms more naturally than their dS counterparts.

\subsection{Finite-order truncations and the bounce}

A natural question is whether the occurrence of a bounce in the full AdS-core solution can be inferred from a finite-order effective description valid in the pre-bounce region. To investigate this issue, we introduce the variable $\mathcal{X} = 2 m \ell^2 / r^3$, allowing the metric function in Eq.~\eqref{eq:fads} to be expressed as
\begin{align}
	f_{\mathrm{AdS}}(\mathcal{X})=1-\frac{2m}{r}(1-\epsilon \mathcal{X})\frac{1}{(1+\mathcal{X})(1+\epsilon\mathcal{X})}. 
\end{align} 
For simplicity, and in analogy with the previous part of this section, we set $\epsilon = 1$. In this case, the expression reduces to
\begin{align}
	f_{\mathrm{AdS}}(\mathcal{X})=1-\frac{2m}{r}(1-\mathcal{X})\frac{1}{(1+\mathcal{X})^2}. 
\end{align}
We now assume that $\mathcal X<1$, corresponding to the exterior of the MS-mass sign-change radius $r_0=(2m\ell^2)^{1/3}$, which coincides with the bounce radius in the marginally bound case. For concreteness, one may consider, for instance, the marginally bound case with $\ell = 0.2$ shown in Fig.~\ref{fig:dotR}, where a bounce indeed occurs. In the following we use a partial resummation scheme in which the factor $(1-\mathcal{X})$ is kept exact, while the denominator is expanded in the exterior region $\mathcal{X}<1$. Under this assumption, the fraction $1/(1+\mathcal{X})$ can be expanded in a convergent power series as
\begin{align}
	\frac{1}{1+\mathcal{X}}=\sum_{n=0}^{\infty}(-1)^{n}\mathcal{X}^n.
\end{align}
Differentiating the above series expansion term by term yields the corresponding expansion for the squared denominator, namely
\begin{align}
	\frac{1}{(1+\mathcal{X})^2}
	= \sum_{n=0}^{\infty} (-1)^n (n+1)\mathcal{X}^n .
\end{align}
We are now interested in a finite truncation of this expansion at order $N$. To derive the truncated expression, we proceed as before and differentiate the truncated series for $1/(1+\mathcal{X})$, which reads
\begin{align}
	\left.\frac{1}{1+\mathcal{X}}\right|_{N}
	= \sum_{n=0}^{N} (-1)^n \mathcal{X}^n
	= \frac{1-(-\mathcal{X})^{N+1}}{1+\mathcal{X}} .
\end{align}
This leads to the truncated expression for $1/(1+\mathcal{X})^2$, which we denote by $S_N(\mathcal{X})$, given by
\begin{align}
	S_N(\mathcal{X})
	= \frac{1 - (-\mathcal{X})^{N+1}\left[(N+2)+(N+1)\mathcal{X}\right]}
	{(1+\mathcal{X})^2}.
\end{align}
\begin{figure}[!htbp]
	\centering
	\includegraphics[scale=0.4]{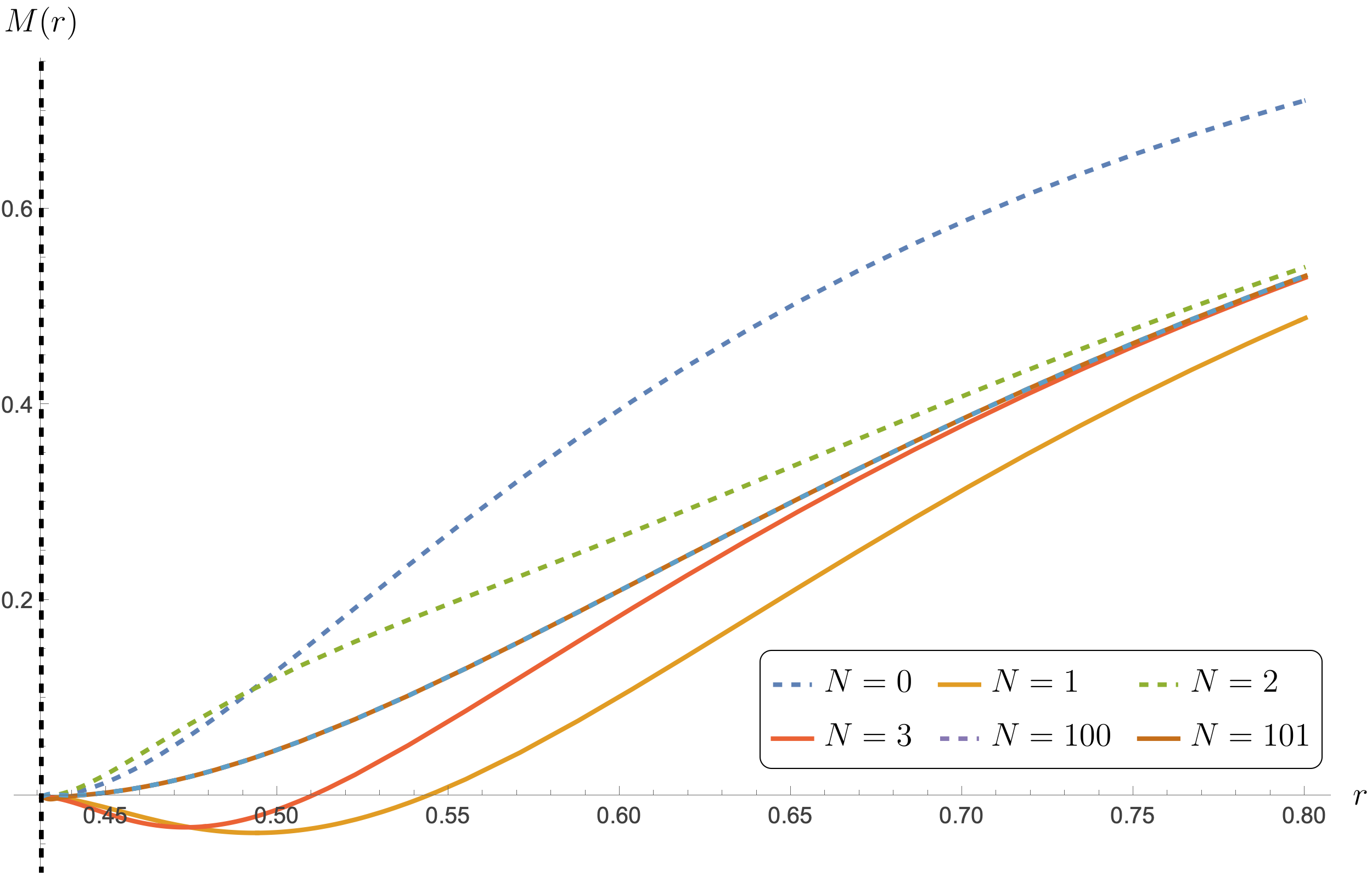}
	\caption{\textbf{Finite $N$ truncations.} The MS mass of the Hayward model with AdS core is plotted for finite $N$ truncations as a function of the radius $r$, as defined in Eq.~\eqref{eq:MS-N}. The mass, the minimal length scale, and the parameter $\epsilon$ are set to $m=1$, $\ell=0.2$, and $\epsilon=1$, respectively, corresponding to a black hole configuration that exhibits a bounce (see Figs.~\ref{fig:f} and \ref{fig:dotR}). The dashed vertical line indicates $r_0=(2m\ell^2)^{1/3}\equiv 0.4309$, namely the MS-mass sign-change radius; for the marginally bound shell used in this illustrative plot, this radius coincides with the bounce radius of the full AdS-core Hayward model. We observe that for sufficiently large $N$, the truncated curves can be nearly indistinguishable from the full solution over a pre-bounce exterior region, even though different finite-order truncations may correspond to different effective continuations toward the core.
	}\label{fig:plotN}
\end{figure}

One can readily show that this expression is always positive for even $N$ in the interval $0<\mathcal{X}<1$. By contrast, for odd $N$ a sign change necessarily occurs. Indeed, since $S_N$ is a continuous function of $\mathcal{X}$, with $S_{N}(0)=1>0$ and $S_{N}(1)=-(N+1)/2<0$, it follows from the intermediate value theorem that there exists a $\mathcal{X}_0\in (0,1)$ such that $S_{N}(\mathcal{X}_0)=0$. This implies a sign change of the truncated expression. We emphasize that this sign change directly translates into a sign change of the MS mass and is therefore intimately connected to the occurrence of a bounce, the marginally bound case providing the clearest illustration. For completeness, the MS mass corresponding to the truncated solutions is given by
\begin{align}
	M(\mathcal{X})=m(1-\mathcal{X})S_{N}(\mathcal{X}).\label{eq:MS-N}
\end{align}

Since $1-\mathcal{X}>0$ in the region under consideration, the sign of the truncated MS mass is controlled by $S_N(\mathcal{X})$. Thus odd and even truncations can lead to qualitatively different finite-order effective models: odd truncations can develop a zero of the MS mass within $0<\mathcal{X}<1$, whereas even truncations do not\footnote{The same qualitative conclusion is obtained if one instead
	truncates the full exterior Taylor expansion
	\[
	\frac{1-\mathcal{X}}{(1+\mathcal{X})^2}
	=
	\sum_{n=0}^{\infty}(-1)^n(2n+1)\mathcal{X}^n,
	\qquad |\mathcal{X}|<1 .
	\]
	For the finite truncation
	$S_N^{\rm full}(\mathcal{X})=\sum_{n=0}^{N}(-1)^n(2n+1)\mathcal{X}^n$,
	one finds that even truncations remain positive on
	$0<\mathcal{X}<1$, whereas odd truncations change sign in this
	interval. Thus the odd/even distinction is not an artifact of
	keeping the numerator exact.}. This should not be interpreted as a failure of the exact theory to predict its own bounce. Rather, it shows that finite-order truncations, although they may approximate the exterior geometry well, need not encode the same inner continuation as the full nonperturbative model.

For sufficiently large $N$, the odd and even truncations can agree to high accuracy in the region $\mathcal{X}<\mathcal{X}_0$, or equivalently for $r>(2m\ell^2/\mathcal{X}_0)^{1/3}$. Therefore, data restricted to a pre-bounce exterior region may be compatible with distinct finite-order effective models, some of which contain an inner turning point and some of which do not. This behavior is illustrated in Fig.~\ref{fig:plotN}. An analogous analysis for the Hayward model with a dS core leads to the same lesson \cite{GLSW:25}. The non-marginally bound cases are more involved, as one must additionally account for the value of $\mathcal{E}$; nevertheless, the same qualitative issue is expected to persist.

Finally, we emphasize that although finite-order exterior truncations may not be sufficient to determine the inner continuation uniquely, the two core geometries themselves can always be distinguished by other means, as they correspond to fundamentally different models. For instance, they can be differentiated through their quasinormal mode spectra \cite{ALNV:25}. Moreover, we note that the $N=1$ truncation for the dS core case reproduces exactly a standard loop quantum gravity–inspired model \cite{APS:06-Letter, APS:06}, given by
\begin{align}
	f_{\mathrm{LQG}}(r)=1-\frac{2m}{r}+\frac{4m^2\alpha^2_{\Delta}}{r^4},
\end{align}
where $\alpha_{\Delta}$ is a parameter related to the Barbero–Immirzi parameter $\gamma$ and the Planck scale $\ell_{p}$ through $\alpha_{\Delta}=\gamma\sqrt{\Delta}$, with $\Delta=4\pi\sqrt{3}\gamma\ell_{p}$. A direct correspondence between the two descriptions can be established via the identification $\alpha_{\Delta}=\ell$ \cite{GLSW:25}. A similar observation applies to the AdS-core Hayward model: in this case, the truncation at $N=0$ yields an analogous structure under the same parameter identification. Since $S_{0}(\mathcal{X})=1$, it follows that
\begin{align}
	f_{\mathrm{AdS}}(r)\big|_{N=0}=1-\frac{2m}{r}+\frac{4m^2\ell^2}{r^4}.
\end{align}
A direct analysis of the possibility of a bounce and of its visibility in finite-order truncations has been carried out in this section. We now turn to the explicit construction of the effective quantum gravity framework, beginning with the computation of the physical Hamiltonian and subsequently deriving the full covariant four-dimensional Lagrangian.

\section{Effective quantum gravity framework}

Any consistent theory of gravity must respect invariance under spatial coordinate transformations, namely spatial diffeomorphism invariance. Within the Hamiltonian framework, this symmetry is implemented through the vector constraint. From the Lagrangian perspective, it requires that the Lagrangian density be constructed exclusively from scalar invariants derived from the intrinsic and extrinsic geometry of spatial hypersurfaces. In the case of spherical symmetry, the available geometric structures are severely restricted, leading to a correspondingly constrained form of the theory. 

In this setting, the relevant curvature invariants are directly encoded in the quantities $\mathcal{K}_1$, $\mathcal{K}_2$, $\mathcal{R}_1$, and $\mathcal{R}_2$. Consequently, the spatially covariant Lagrangian can depend only on these variables. Moreover, there exists a straightforward procedure to derive it directly from the modified scalar constraint \cite{GL:25}. In this section, we first identify the Hamiltonian corresponding to the Hayward model with an AdS core and analyze some of its key properties. We then use this structure to construct the associated spatially covariant Lagrangian, and finally promote it to a fully covariant four-dimensional Lagrangian.

\subsection{Derivation of the physical Hamiltonian}\label{sec:Hamiltonian}

\begin{figure*}[!htbp]
	\centering
	{
		\includegraphics[scale=0.39]{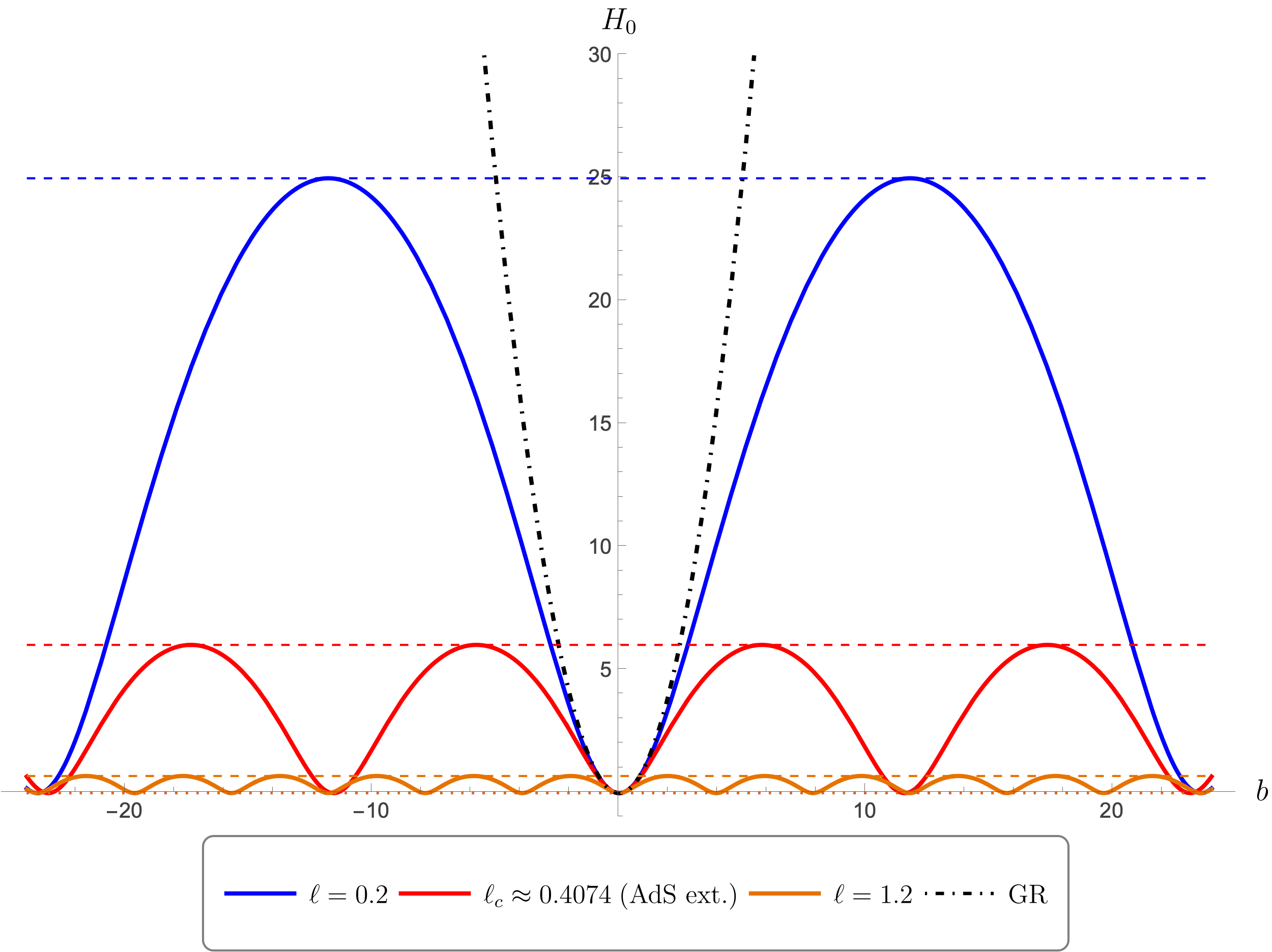}\
	}
	\hspace{0.05\textwidth}
	{
		\includegraphics[scale=0.39]{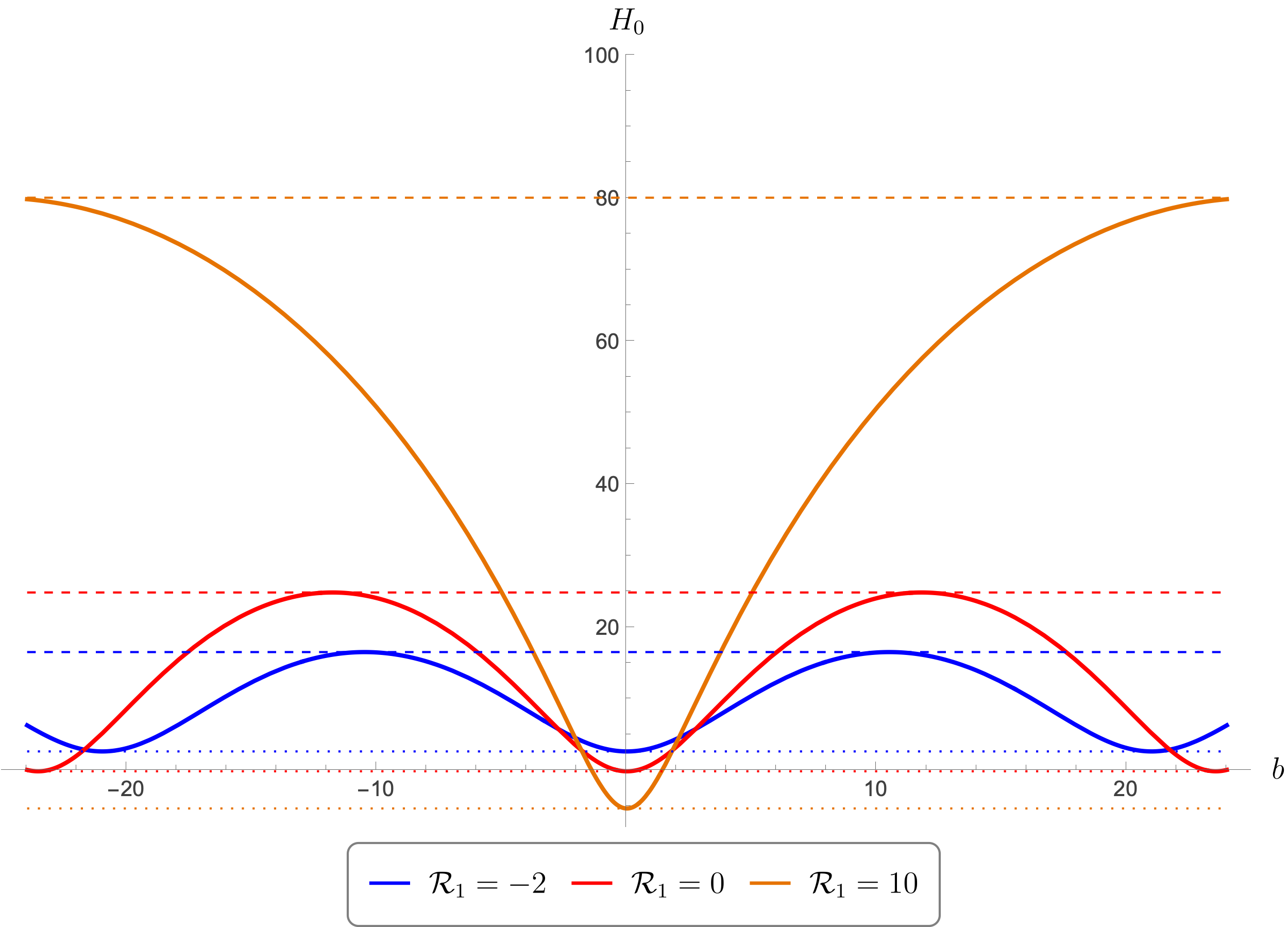}
	}
	\caption{\textbf{Low-curvature branch of the physical Hamiltonian.}
		The function $H_0$ is shown as a function of the canonical variable $b$, with the integration function fixed to $b_0(\mathcal{R}_1)=0$. In this branch the Hamiltonian is bounded between the two roots $H_-\leq H_0\leq H_+$; the horizontal dashed and dotted lines indicate the corresponding upper and lower root values, respectively. The black dot-dashed curve in the left panel shows the GR limit for comparison. \textbf{Left:} Varying the minimal length scale $\ell$ at fixed $\mathcal{R}_1=0$. \textbf{Right:} Varying the curvature parameter $\mathcal{R}_1$ at fixed $\ell=0.2$, with all choices lying in the low-curvature domain $-1/(8\ell^2)\leq \mathcal{R}_1<1/\ell^2$.}
	\label{fig:plotH0-lowcurvature}
\end{figure*}

\begin{figure*}[!htbp]
	\centering
	{
		\includegraphics[scale=0.39]{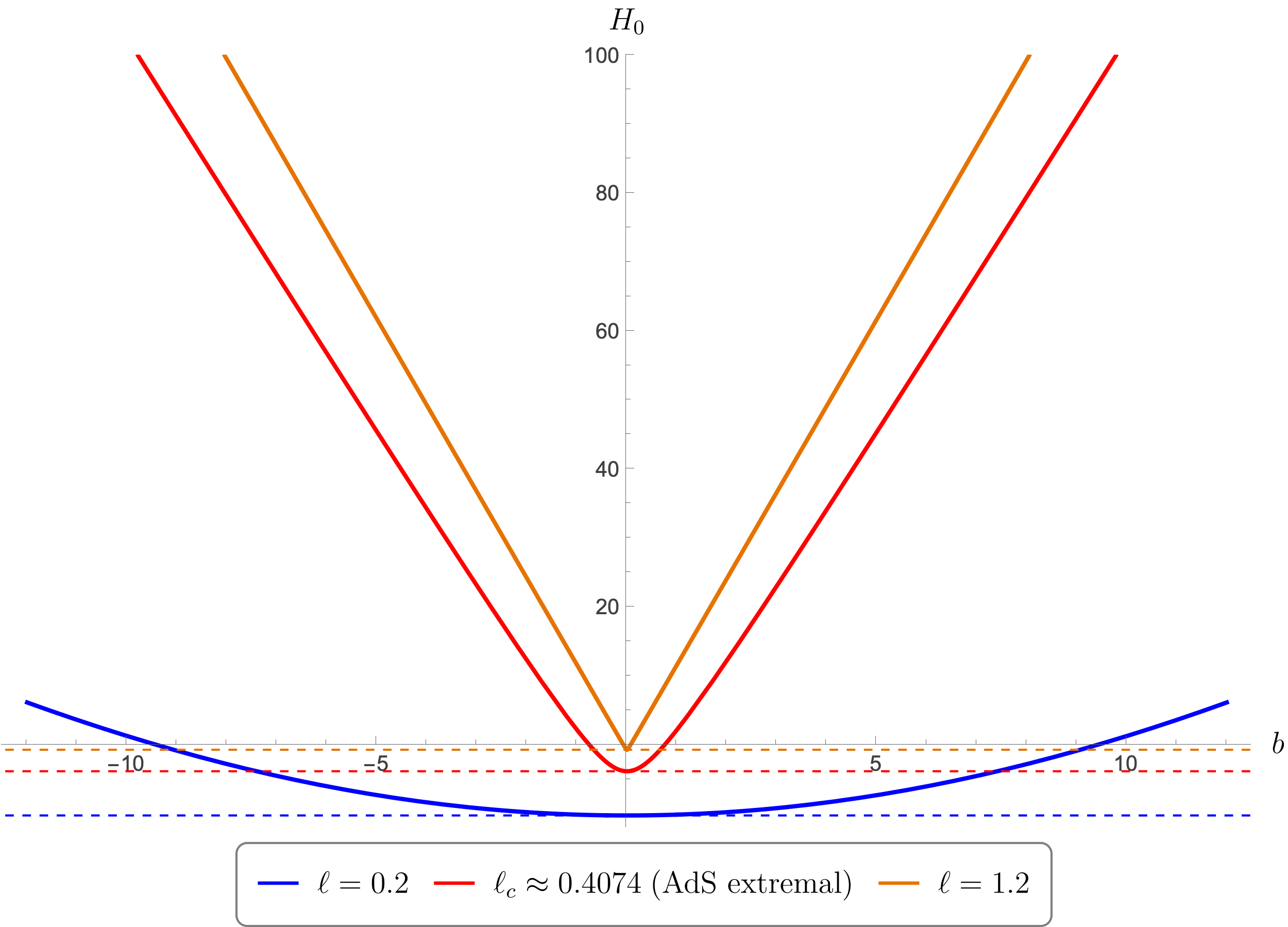}\
	}
	\hspace{0.05\textwidth}
	{
		\includegraphics[scale=0.39]{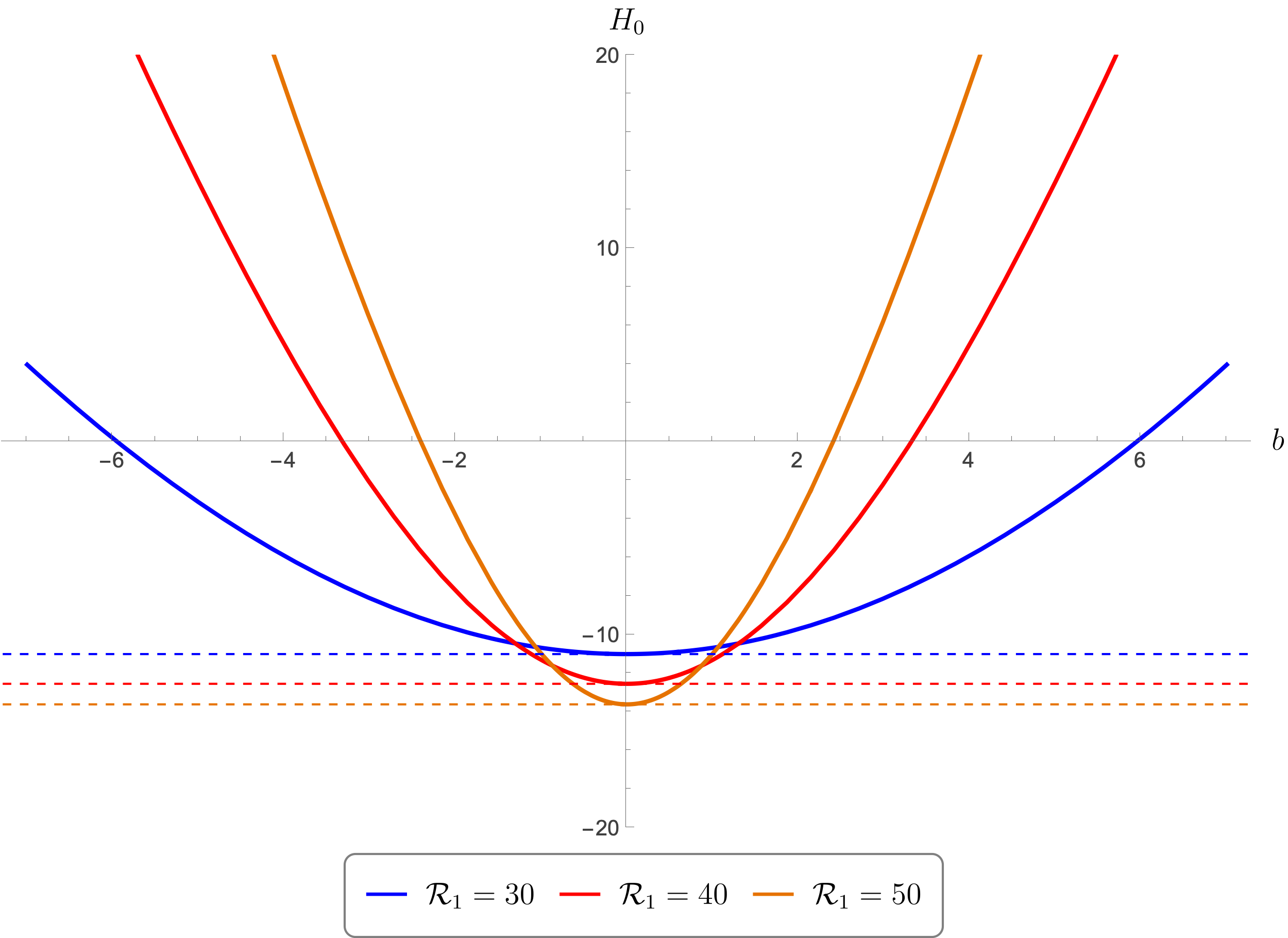}
	}
	\caption{\textbf{High-curvature branch of the physical Hamiltonian.}
		The function \(H_0\) is shown as a function of the canonical variable \(b\), with the integration function fixed to \(b_0(\mathcal{R}_1)=0\). This branch applies for \(\ell^2\mathcal{R}_1>1\). The horizontal dashed lines indicate the corresponding lower bound \(H_>\), and the solution satisfies \(H_0\geq H_>\) for all \(b\). \textbf{Left:} Varying the minimal length scale \(\ell\) at fixed \(\mathcal{R}_1=40\), with the displayed values chosen in the high-curvature domain. \textbf{Right:} Varying the curvature parameter \(\mathcal{R}_1\) at fixed \(\ell=0.2\), again with all choices satisfying \(\ell^2\mathcal{R}_1>1\).}
	\label{fig:plotH0-highcurvature}
	
\end{figure*}

Determining the physical Hamiltonian reduces to solving a partial differential equation of the form
\begin{align}
	\left(\partial_b H_0\right)^2 = 4\left(\mathcal{R}_1+\tilde f(H_0)\right).\label{eq:PDE-simple}
\end{align}
For the AdS-core Hayward model considered in this paper, and after setting \(\epsilon=1\), this becomes
\begin{align}
	\left(\partial_b H_0\right)^2
	=
	4\left[
	\mathcal{R}_1+
	\frac{H_0(1-\ell^2 H_0)}{(1+\ell^2 H_0)^2}
	\right].
	\label{eq:H0-PDE-eps1}
\end{align}
The choice $\epsilon=1$ is special because the denominator becomes a perfect square. For generic $\epsilon$, the corresponding integral is expressed in terms of elliptic integrals, as discussed in Appendix~\ref{sec:app:Elliptic}. For $\epsilon=1$, however, the problem admits a more explicit branchwise representation.

Let us introduce
\begin{align}
	c_\ell\equiv \frac{1}{\ell^2},
	\qquad
	x\equiv \ell^2 \mathcal{R}_1,
	\qquad
	\Delta\equiv \sqrt{1+8x}.
\end{align}
For $x\neq 1$, Eq.~\eqref{eq:H0-PDE-eps1} can be written as
\begin{align}
	\left(\partial_b H_0\right)^2
	=
	\frac{
		4(\mathcal{R}_1-c_\ell)(H_0-H_<)(H_0-H_>)
	}{(H_0+c_\ell)^2},
	\label{eq:H0-factorized}
\end{align}
where \(H_<\) and \(H_>\) denote the smaller and larger roots of the numerator. Equivalently, the two roots are
\begin{align}
	H_{\pm}
	=
	\frac{1+2x\pm \Delta}{2\ell^2(1-x)},
\end{align}
with the ordering exchanged when crossing \(x=1\). The roots are real provided
\begin{align}
	\mathcal{R}_1\geq -\frac{1}{8\ell^2}.
\end{align}
Thus the explicit Hamiltonian naturally separates into different branches.

First, we consider the low-curvature branch,
\begin{align}
	-\frac{1}{8\ell^2}\leq \mathcal{R}_1<c_\ell .
\end{align}
In this regime $\mathcal{R}_1-c_\ell<0$, so the physical solution lies between the two roots. We therefore write
\begin{align}
	H_0
	=
	H_<+(H_>-H_<)\sin^2\frac{\theta}{2},
	\qquad
	0\leq \theta\leq \pi .
\end{align}
Substitution into Eq.~\eqref{eq:H0-factorized} gives
\begin{align}
	\lambda_{\rm K}\theta-\sin\theta
	=
	\pm
	\frac{4\sqrt{c_\ell-\mathcal{R}_1}}{H_>-H_<}
	\left[b-b_0(\mathcal{R}_1)\right],
	\label{eq:kepler-branch}
\end{align}
where \(b_0(\mathcal{R}_1)\) is an integration function corresponding to the freedom to shift the canonical variable \(b\), and
\begin{align}
	\lambda_{\rm K}
	=
	\frac{H_<+H_>+2c_\ell}{H_>-H_<}
	=
	\frac{3}{\sqrt{1+8\ell^2\mathcal{R}_1}} .
\end{align}
Since \(\lambda_{\rm K}\geq 1\) throughout this branch, the map \(\lambda_{\rm K}\theta-\sin\theta\) is monotonic and can be inverted uniquely.\footnote{After division by \(\lambda_{\rm K}\), Eq.~\eqref{eq:kepler-branch} takes the form \(\theta-e\sin\theta=\Xi\), with \(e=1/\lambda_{\rm K}<1\) away from the critical surface. This is the standard Kepler equation. Here this identification is used only to characterize the inverse map; no orbital interpretation is implied.} In particular, the marginally bound case \(\mathcal{R}_1=0\) lies in this branch. There one obtains
\begin{align}
	H_0=\frac{1}{\ell^2}\sin^2\frac{\theta}{2},
	\qquad
	3\theta-\sin\theta
	=
	\pm 4\ell\left[b-b_0(0)\right],
\end{align}
so that \(0\leq H_0\leq 1/\ell^2\). Characteristic examples of this branch, including the marginally bound case, are shown in Fig.~\ref{fig:plotH0-lowcurvature}.

We now turn to the high-curvature branch,
\begin{align}
	\mathcal{R}_1>c_\ell .
\end{align}
In this regime \(\mathcal{R}_1-c_\ell>0\), and the positive-mass branch relevant for the present construction is described by \(H_0\geq H_>\). We set
\begin{align}
	H_0
	=
	H_>+(H_>-H_<)\sinh^2 y .
\end{align}
Eq.~\eqref{eq:H0-factorized} then integrates to
\begin{align}
	\sinh(2y)+\lambda_{\rm L}(2y)
	=
	\pm
	\frac{4\sqrt{\mathcal{R}_1-c_\ell}}{H_>-H_<}
	\left[b-b_0(\mathcal{R}_1)\right],
	\label{eq:leal-branch}
\end{align}
where
\begin{align}
	\lambda_{\rm L}
	=
	\frac{H_<+H_>+2c_\ell}{H_>-H_<}
	=
	-\frac{3}{\sqrt{1+8\ell^2\mathcal{R}_1}} .
\end{align}
For \(\mathcal{R}_1>c_\ell\), one has \(-1<\lambda_{\rm L}<0\). Therefore the monotonicity condition derived in Appendix~\ref{sec:app:Leal} applies directly. Using the notation of Appendix~\ref{sec:app:Leal}, Eq.~\eqref{eq:leal-branch} is inverted as
\begin{align}
	y
	=
	\frac{1}{2}
	L_{\lambda_{\rm L}}{\sinh_2}
	\left[
	\pm
	\frac{4\sqrt{\mathcal{R}_1-c_\ell}}{H_>-H_<}
	\left(b-b_0(\mathcal{R}_1)\right)
	\right],
\end{align}
where \(L_{\lambda}{\sinh_2}\) denotes the inverse of \(z\mapsto \sinh z+\lambda z\), i.e., the generalized Leal function. Hence the high-curvature branch of the Hamiltonian is
\begin{widetext}
	\begin{align}
		H_0(b,R_1)
		=
		H_>+(H_>-H_<)
		\sinh^2
		\left\{
		\frac{1}{2}
		L_{\lambda_{\rm L}}{\sinh_2}
		\left[
		\pm
		\frac{4\sqrt{\mathcal{R}_1-c_\ell}}{H_>-H_<}
		\left(b-b_0(\mathcal{R}_1)\right)
		\right]
		\right\}.
		\label{eq:H0-high-branch}
	\end{align}
\end{widetext}
This branch is illustrated in Fig.~\ref{fig:plotH0-highcurvature} for different values of the curvature \(\mathcal{R}_1\) and the minimal length scale $\ell$. The oscillatory structure in the low-curvature branch reflects the bounded trigonometric parametrization of \(H_0(b,\mathcal{R}_1)\) when it is extended over several branches of the canonical variable \(b\). This is analogous to bounded polymerization functions in loop-inspired models \cite{GLSW:25}, where the full periodic function is defined on a cover of phase space but a physical reconstruction is performed on monotonic segments. Thus the repeated lobes in the plot should not be interpreted as an oscillatory evolution of the shell radius; they display the branch structure of the map from \(b\) to \(H_0\).

The surface \(\mathcal{R}_1=c_\ell\), or \(x=1\), separates the two branches. At this point the quadratic coefficient of the numerator in Eq.~\eqref{eq:H0-PDE-eps1} vanishes, one of the roots is sent to infinity, and the two-root parametrization degenerates. The differential equation reduces directly to
\begin{align}
	\left(\partial_b H_0\right)^2
	=
	4\frac{c_\ell+3H_0}{(1+\ell^2H_0)^2}.
\end{align}
This should be understood as the limiting case connecting the trigonometric and hyperbolic representations.

The GR limit is most cleanly taken directly at the level of Eq.~\eqref{eq:H0-PDE-eps1}. Sending \(\ell\to0\), one obtains
\begin{align}
	\left(\partial_b H_0\right)^2
	=
	4(\mathcal{R}_1+H_0),
\end{align}
whose solution is
\begin{align}
	H_0^{\rm GR}
	=
	\left[b-b_0(\mathcal{R}_1)\right]^2-\mathcal{R}_1 .
\end{align}
This is the expected GR shell Hamiltonian, up to the irrelevant shift \(b_0(\mathcal{R}_1)\).

We therefore obtain a branchwise analytic representation of the physical Hamiltonian,
\begin{align}
	H=vH_0(b,\mathcal{R}_1),
\end{align}
for the \(\epsilon=1\) AdS-core Hayward model. The low-curvature branch contains the marginally bound and asymptotically flat regimes, whereas the high-curvature branch is described by the generalized Leal function introduced in Appendix~\ref{sec:app:Leal}. For non-marginally unbound shells, \(\mathcal{R}_1=\mathcal{E}/r^2>0\), an infalling trajectory may pass from the low-curvature branch to the high-curvature branch across \(\mathcal{R}_1=1/\ell^2\). For bound shells, \(\mathcal{R}_1=\mathcal{E}/r^2<0\), the reality condition becomes \(r\geq r_{\rm H}\equiv \ell\sqrt{-8\mathcal E}\). This is a domain condition for the explicit Hamiltonian representation, not a bounce condition; in the bounded AdS-core Hayward case the turning points, whenever present, remain within this real domain \footnote{For the \(\epsilon=1\) AdS-core Hayward mass function,
	\[
	M(r)=\frac{m r^3(r^3-2m\ell^2)}{(r^3+2m\ell^2)^2},
	\]
	this compatibility can be checked explicitly in the bounded case. Writing \(\mathcal E=-q<0\) and \(y=r^3/(2m\ell^2)\), the turning-point equation \(\mathcal E+2M(r)/r=0\) becomes
	\[
	q=\left(\frac{2m}{\ell}\right)^{2/3}
	\frac{y^{2/3}(y-1)}{(y+1)^2}.
	\]
	Thus any bounded-shell turning point has \(y>1\). Moreover, using the same equation one finds
	\[
	r_{\rm turn}^2-r_{\rm H}^2
	=
	r_{\rm turn}^2\frac{(y-3)^2}{(y+1)^2}\geq0,
	\qquad
	r_{\rm H}\equiv \ell\sqrt{-8\mathcal E}.
	\]
	Hence, whenever bounded turning points exist, they lie in the real domain of the explicit Hamiltonian representation.}.
These branch domains characterize the explicit Hamiltonian representation and should not be confused with the bounce criteria studied in Sec.~\ref{sec:bounce}, which are determined by the zeros of \(\mathcal{E}+2M(r)/r\).

Finally, we stress that the sign change of the local Misner--Sharp mass in the AdS-core geometry should not be confused with a sign change of the conserved shell Hamiltonian. On the positive-mass static vacuum branch one has
\begin{align}
	H=vH_0=r^3H_0=2m>0,
\end{align}
or equivalently
\begin{align}
	H_0=\frac{2m}{r^3}>0.
\end{align}

Thus the conserved shell Hamiltonian $H$, or equivalently the mass parameter $m$, remains positive throughout the vacuum reconstruction, while $H_0$ is the local argument entering the function $\tilde f$. By contrast, the local Misner--Sharp mass of the static geometry is
\begin{align}
	M_{\rm stat}(r)
	=
	\frac{r^3}{2}\tilde f\left(\frac{2m}{r^3}\right),
\end{align}
and may change sign because the modified function \(\tilde f\) changes sign. Thus the negative effective Misner--Sharp mass near the AdS core does not require \(H_0\) itself to become negative. Negative values appearing in extended branch plots of \(H_0(b,\mathcal{R}_1)\), if present, should therefore not be identified with the AdS-core mass sign change.

\subsection{Derivation of the four-dimensional covariant Lagrangian}\label{sec:Lagrangian}
Following the procedure of Ref.~\cite{GL:25}, one is naturally led to a spatially covariant Lagrangian expressed in terms of the intrinsic and extrinsic curvature invariants. In the conventions used here, it is given by
\begin{widetext}
	\begin{align}
		\mathcal{L}
		=
		\sqrt{h}\Bigg[
		\frac{\mathcal{B}}{2}
		\left(\mathcal{K}_{2}+2\mathcal{K}_{1}\right)
		-3H_{0}
		+2\left(\mathcal{R}_{1}+\mathcal{R}_{2}\right)
		\partial_{\mathcal{R}_{1}}H_{0}
		-\mathcal{K}_{1}
		\left(\frac{\mathcal{K}_{2}}{2}-2\mathcal{K}_{1}\right)
		\partial_{\mathcal{K}_{1}}\mathcal{B}
		+
		2\mathcal{K}_{1}
		\left(\mathcal{R}_{1}+\mathcal{R}_{2}\right)
		\partial_{\mathcal{R}_{1}}\mathcal{B}
		\Bigg].
		\label{eq:Lsp}
	\end{align}
\end{widetext}

Here \(\mathcal{B}=\mathcal{B}(\mathcal{R}_{1},\mathcal{K}_{1})\) denotes the local inverse that expresses the canonical variable \(b\) in terms of the curvature variables,
\begin{align}
	b=\mathcal{B}(\mathcal{R}_{1},\mathcal{K}_{1}).
\end{align}
It is defined implicitly by
\begin{align}
	\mathcal{K}_{1}
	=
	\frac{1}{2}\partial_b H_0(b,\mathcal{R}_{1}),
	\label{eq:K1-dH0}
\end{align}
as derived below. In Eq.~\eqref{eq:Lsp}, all appearances of \(H_0\) and \(\partial_{\mathcal{R}_{1}}H_0\) are evaluated at \(b=\mathcal{B}(\mathcal{R}_{1},\mathcal{K}_{1})\), with \(\partial_{\mathcal{R}_{1}}H_0\) taken at fixed \(b\) before this substitution. The derivatives of \(\mathcal B\) are taken with respect to its displayed arguments \((\mathcal{R}_{1},\mathcal{K}_{1})\).

To see where Eq.~\eqref{eq:K1-dH0} comes from, start from the Hamiltonian \(H=vH_0\), with \(v=(E^x)^{3/2}\). Hamilton's equation gives
\begin{align}
	\dot v=\frac{3}{2}\partial_b H.
\end{align}
On the other hand, using the definition of \(\mathcal{K}_1\), one has
\begin{align}
	\dot v=3v\mathcal{K}_1.
\end{align}
Combining these two expressions yields precisely Eq.~\eqref{eq:K1-dH0}.

Since the Hamiltonian found in Sec.~\ref{sec:Hamiltonian} is branchwise, the inverse function \(\mathcal{B}\) is branchwise as well. On the high-curvature branch, \(\mathcal{R}_1>c_\ell\), we write
\begin{align}
	H_0=H_>+(H_>-H_<)\sinh^2 y ,
\end{align}
where \(H_>\) and \(H_<\) denote the larger and smaller roots, respectively. Defining \(Y=2y\), Eq.~\eqref{eq:K1-dH0} gives
\begin{align}
	\mathcal{K}_1
	=
	\pm
	\frac{\sqrt{\mathcal{R}_1-c_\ell}\sinh Y}
	{\lambda_{\rm L}+\cosh Y}.
	\label{eq:K1-high}
\end{align}
Solving this algebraic relation gives a local inverse \(Y=Y(\mathcal{R}_1,\mathcal{K}_1)\). Equivalently, one may write
\begin{align}
	\mathcal{B}_{\rm L}(\mathcal{R}_1,\mathcal{K}_1)
	=
	b_0(\mathcal{R}_1)
	\pm
	\frac{H_>-H_<}{4\sqrt{\mathcal{R}_1-c_\ell}}
	\left(\sinh Y+\lambda_{\rm L}Y\right),
	\label{eq:B-high}
\end{align}
where the sign labels the local branch of the inverse map. This is precisely the branch controlled by the generalized \(\lambda\)-Leal function introduced in Appendix~\ref{sec:app:Leal}.

On the low-curvature branch,
\begin{align}
	-\frac{1}{8\ell^2}\leq \mathcal{R}_1<c_\ell ,
\end{align}
the Hamiltonian is instead written in trigonometric form,
\begin{align}
	H_0
	=
	H_<+(H_>-H_<)\sin^2\frac{\theta}{2}.
\end{align}
In this case Eq.~\eqref{eq:K1-dH0} becomes
\begin{align}
	\mathcal{K}_1
	=
	\pm
	\frac{\sqrt{c_\ell-\mathcal{R}_1}\sin\theta}
	{\lambda_{\rm K}-\cos\theta}.
	\label{eq:K1-low}
\end{align}
Thus the corresponding inverse \(\theta=\theta(\mathcal{R}_1,\mathcal{K}_1)\) determines
\begin{align}
	\mathcal{B}_{\rm K}(\mathcal{R}_1,\mathcal{K}_1)
	=
	b_0(\mathcal{R}_1)
	\pm
	\frac{H_>-H_<}{4\sqrt{c_\ell-\mathcal{R}_1}}
	\left(\lambda_{\rm K}\theta-\sin\theta\right).
	\label{eq:B-low}
\end{align}
This low-curvature branch includes the marginally bound and asymptotically flat regimes. A global covariant completion should therefore be understood branchwise, with the Leal representation used for \(\mathcal{R}_1>c_\ell\), the trigonometric representation used for \(\mathcal{R}_1<c_\ell\), and the critical surface \(\mathcal{R}_1=c_\ell\) treated as the limiting case connecting the two. The Lagrangian should therefore be evaluated by taking the limiting form of the inverse map from either side. 

The apparent singularity of the inverse maps at
$\mathcal R_1=c_\ell$ is removable. In Appendix~\ref{sec:app:branch-matching}
we show explicitly that $\mathcal B_{\rm K}$ and $\mathcal B_{\rm L}$ have
the same limiting value at this surface, and that the derivatives
$\partial_{\mathcal K_1}\mathcal B$ and
$\partial_{\mathcal R_1}\mathcal B$ entering Eq.~\eqref{eq:Lsp}
also admit regular one-sided limits, provided $b_0(\mathcal R_1)$ is
chosen smoothly and the local inverse is nondegenerate. Thus
$\mathcal R_1=c_\ell$ does not introduce an additional branch
singularity in the covariant Lagrangian.

The next step is to express \(\mathcal{K}_1\), \(\mathcal{K}_2\), \(\mathcal{R}_1\), and \(\mathcal{R}_2\) in terms of scalar invariants. For the extrinsic curvature we have
\begin{align}
	\tensor{K}{^i_i}
	=
	\mathcal{K}_1+\mathcal{K}_2,
	\qquad
	K_{ij}K^{ij}
	=
	3\mathcal{K}_1^2+\mathcal{K}_2^2
	-2\mathcal{K}_1\mathcal{K}_2 .
\end{align}
Solving these relations gives
\begin{align}
	\mathcal{K}_1
	=
	\frac{1}{6}
	\left(
	2\tensor{K}{^i_i}
	\mp
	\sqrt{
		2\left(3K_{ij}K^{ij}-(\tensor{K}{^i_i})^2\right)}
	\right),
\end{align}
and
\begin{align}
	\mathcal{K}_2
	=
	\frac{1}{6}
	\left(
	4\tensor{K}{^i_i}
	\pm
	\sqrt{
		2\left(3K_{ij}K^{ij}-(\tensor{K}{^i_i})^2\right)}
	\right).
\end{align}
The sign choice here should be understood as an eigenvalue-labeling prescription: it identifies which eigenvalue corresponds to the angular direction used in the spherical reduction. It is not an independent change of theory between collapsing and expanding phases. Along a bouncing solution the same eigenvalue label should be continued smoothly through the turning point.

Similarly, with our convention for \(\mathcal{R}_1\), the intrinsic curvature invariants are
\begin{align}
	{}^{(3)}R
	=
	-2\mathcal{R}_1+4\mathcal{R}_2,
	\qquad
	{}^{(3)}\mathcal{K}
	=
	4\mathcal{R}_1^2+8\mathcal{R}_2^2 .
	\label{eq:R-scalar-relations}
\end{align}
Solving for \(\mathcal{R}_1\) and \(\mathcal{R}_2\), we obtain
\begin{align}
	\mathcal{R}_1
	=
	\frac{1}{6}
	\left(
	-{}^{(3)}R
	\pm
	\sqrt{
		2\left(3{}^{(3)}\mathcal{K}-({}^{(3)}R)^2\right)}
	\right),
\end{align}
and
\begin{align}
	\mathcal{R}_2
	=
	\frac{1}{12}
	\left(
	2{}^{(3)}R
	\pm
	\sqrt{
		2\left(3{}^{(3)}\mathcal{K}-({}^{(3)}R)^2\right)}
	\right),
\end{align}
with correlated signs. On the LTB vacuum sector, where \(\mathcal{R}_2=0\), the plus sign corresponds to \(\mathcal{R}_1>0\), while the minus sign corresponds to \(\mathcal{R}_1<0\). More generally, this sign ambiguity is again an eigenvalue-labeling ambiguity of the covariant lift. Away from exact spherical symmetry, a fully off-shell extension would require an explicit projector or branch prescription selecting the appropriate eigenvalue; otherwise the result should be interpreted as a covariant completion of the spherical/eigenvalue sector.

Finally, the spatially covariant Lagrangian can be promoted to a four-dimensional covariant form by introducing a constrained scalar reference field \(\phi\). The unit-timelike condition on \(\phi\) defines the normal to the preferred foliation, and through the replacement
\begin{align}
	\mathcal{P}(K_{ab})=\nabla_a\nabla_b\phi
\end{align}
the extrinsic-curvature dependence of the reconstructed spherical sector is written in covariant form. The three-dimensional curvature scalars are then rewritten in terms of four-dimensional quantities using the Gauss--Codazzi relations and the \(1+3\) decomposition summarized in Appendix~\ref{sec:app:curvature-scalars}. In this sense, Eq.~\eqref{eq:Lsp} provides a four-dimensional covariant completion of the reconstructed spherical/eigenvalue sector associated with the selected branch of the effective Hamiltonian. We do not claim that this completion is unique, nor that it defines a unique fully off-shell extension away from the sector used in the reconstruction.

The resulting covariant theory belongs to the class of generalized extended mimetic gravity models \cite{TK:17,LMNV:19}. Since the construction is reconstructed from a deparameterized Hamiltonian system with second-order evolution in the preferred dust time, the higher-curvature terms enter through projected-curvature and extrinsic-curvature combinations rather than through unconstrained higher time derivatives. In the reconstructed deparameterized sector, this structure prevents the appearance of unconstrained higher time derivatives, and therefore avoids the usual Ostrogradsky-type instability associated with such terms. The scalar reference field is auxiliary within this sector: its constraint fixes the foliation, while the dust density is set to zero on the vacuum solutions considered here. Thus, in the polymerized vacuum sector analyzed in this work, the scalar/dust sector should not be counted as an additional physical matter source or as an independent propagating degree of freedom. A fully off-shell statement about the degrees of freedom of the covariant completion would require a separate Hamiltonian constraint analysis.

\section{Discussion}\label{sec:discussion}

We have performed a detailed study of spherically symmetric ultracompact objects whose singularity is resolved through the introduction of an AdS core. Our analysis is motivated by several works emphasizing that, once quantum effects are taken into account, negative energy contributions can arise in the vicinity of the center \cite{ABRG:22,ABRG:24,RT:23,R:23}. We briefly reviewed the construction of regular geometries with an AdS core following Ref.~\cite{ALNV:25}. In particular, we summarized their main energetic properties, including the behavior of the MS mass, the energy density, and the corresponding radial and tangential pressures. We showed that these configurations are characterized by a negative energy density near the center and contrasted this feature with the dS core case. We then turned to the LTB framework of gravitational collapse and exploited the useful decoupling of the shells, which led to a general prescription for generating regular objects with an AdS core as polymerized vacuum solutions. Finally, we developed the corresponding effective quantum gravity framework by explicitly deriving the physical Hamiltonian and the associated covariant Lagrangian. Below, we summarize some of the main findings of this work.

We have presented a consistent procedure for generating spherically symmetric static solutions, including regular geometries with an AdS core, which constitute the main focus of this work. This construction is achieved within the LTB collapse framework by exploiting the decoupling of the shells, leading to the static solution given in Eq.~\eqref{eq:f-LTB}. The resulting expression contains the function $\tilde{f}(2m/r^3)$, which plays a central role in encoding quantum gravitational corrections, as well as other modifications. Importantly, such effects enter in a constrained manner, namely through the combination $2m/r^3$. Moreover, these static solutions satisfy a Birkhoff-type theorem and are uniquely determined by their mass at the effective level.

We analyzed the occurrence of a bounce in the AdS case and compared it with the dS counterpart. In the marginally bound regime, we found that the AdS configuration always exhibits a bounce, which is intrinsically linked to the negativity of the MS mass near the center. The bounce occurs significantly deeper within the core for black hole solutions than for horizonless configurations, a behavior closely related to the spatial extent of the negative MS mass, as illustrated in Figs.~\ref{fig:subfig:MS} and \ref{fig:dotR}. A similar pattern emerges in the unbounded case: the dS configuration does not admit a bounce, whereas the AdS case does, but only within a specific range of parameters $\ell$ or $\mathcal{E}$. This behavior reflects the competition between the repulsive contribution associated with the MS mass and the kinetic energy of the shell. In the bounded regime, both cases can admit two turning points for suitable parameter choices. The outer turning point corresponds to the recollapse of a closed universe configuration. Additionally, we find that the Hayward model with an AdS core exhibits bounce behavior that closely parallels that of the loop quantum gravity–inspired model of Ref.~\cite{GL:25}. This correspondence suggests that AdS cores may more naturally capture certain features of quantum gravity corrections.

Finally, we constructed an effective quantum gravity framework for describing regular AdS-core solutions as polymerized vacuum configurations. We began with the physical Hamiltonian and showed that, for the representative choice $\epsilon=1$, it admits an explicit branchwise analytic representation: the low-curvature sector is described by a trigonometric inverse map, while the high-curvature sector is governed by the generalized $\lambda$-Leal function. Using this Hamiltonian, we derived the corresponding spatially covariant Lagrangian and then promoted it to a four-dimensional covariant completion by introducing a constrained scalar reference field. This scalar should not be interpreted as a physical matter source. On the vacuum branch, the deparameterizing dust density is set to zero; the scalar/dust sector remains only as an auxiliary clock and foliation variable needed to write the deparameterized spatial theory in covariant form. Its derivatives nevertheless enter nontrivially through the projected-curvature and extrinsic-curvature couplings that define the covariant completion. In this sense, the construction belongs to the broader class of generalized extended mimetic gravity models, while avoiding the introduction of an independent propagating scalar degree of freedom within the polymerized vacuum sector considered here.

This framework opens several promising directions for future research. There are two main reasons for this. First, the solutions are constructed within a fully four-dimensional setting, which we regard as the most realistic scenario currently available. Second, by virtue of a Birkhoff-type theorem, the solutions are uniquely characterized by their mass and admit a well-defined effective description. This effective formulation provides a natural starting point for connecting the theory with phenomenology. In particular, gravitational-wave signals and black-hole shadow observations can serve as sensitive probes of the distinctive features predicted by different ultracompact-object models. The same framework also makes it possible to study perturbations of both black-hole and horizonless configurations, including their quasinormal-mode spectra, light-ring structure, and shadow properties. Related questions concerning stability, mass inflation, and inner-extremal black holes were analyzed from this perspective in Ref.~\cite{LS:26-mass-inflation}.

\section*{acknowledgments}
This work is supported by the National Natural Science Foundation of China (Grant No. 12505081) and the start-up funding from Westlake University. I.S. is supported by the Institute for Theoretical Sciences at Westlake University.

\appendix

\section{General integration procedure for the physical Hamiltonian}\label{sec:app:Elliptic}

In this Appendix, we derive the generic solution to the partial differential equation given in Eq.~\eqref{eq:PDE-simple}. We define 
\begin{align}
	c_1=\frac{1}{\ell^2}, \quad c_2=\frac{1}{\epsilon \ell^2}.
\end{align}
A direct analysis yields
\begin{align}
	\partial_{b}H_{0}=\sqrt{4\frac{(\mathcal{R}_1-c_1)(H_0-H_{+})(H_0-H_{-})}{(H_0+c_1)(H_0+c_2)}},
\end{align}
which translates to
\begin{widetext}
	\begin{align}
		\int \left(4\frac{(\mathcal{R}_1-c_1)(H_0-H_{+})(H_0-H_{-})}{(H_0+c_1)(H_0+c_2)}\right)^{-1/2}dH_0=b-b_0(\mathcal{R}_1).
	\end{align}
\end{widetext}

The integral in this form cannot be expressed in terms of elementary functions. However, it admits a compact representation in terms of incomplete elliptic integrals of the first, second, and third kind \cite{H:1910}. Before presenting the explicit result, we briefly recall their definitions. The elliptic integral of the first kind is given by
\begin{align}
	F(\phi,k)=\int_0^{\phi}\frac{d\theta}{\sqrt{1-k^2\sin^2\theta}},
\end{align}  
the second kind by
\begin{align}
	E(\phi,k)=\int_{0}^{\phi}d\theta\sqrt{1-k^2\sin^2\theta},
\end{align}
and the third kind by
\begin{align}
	\Pi(n;\phi,k)=\int_{0}^{\phi}\frac{d\theta}{(1-n\sin^2\theta)\sqrt{1-k^2\sin^2\theta}}.
\end{align}
The solution to the partial differential equation can therefore be written in the compact form
\begin{widetext}
	\begin{align}
		b-b_0(\mathcal{R}_1)=A_0(H_0)\bigg(B_0(H_0)+B_{\mathrm{F}}(H_0)F(\phi,k)+B_{\mathrm{E}}(H_0)E(\phi,k)+B_{\mathrm{\Pi}}(H_0)\Pi(n;\phi,k)\bigg),
	\end{align}
\end{widetext}
where the coefficients $A_0$, $B_0$, $B_{\mathrm{F}}$, $B_{\mathrm{E}}$, and $B_{\mathrm{\Pi}}$, as well as the argument $\phi$, are functions of the physical Hamiltonian $H_0$. The explicit expressions for the coefficients, the argument $\phi$, and the parameters $k$ and $n$ are given below:
\begin{widetext}
	\begin{align}
		A_0(H_0)=\frac{ \sqrt{H_0 - H_{-}} \sqrt{H_0 - H_{+}} }{ 2 \sqrt{c_1 + H_0} (c_2 + H_0) \sqrt{ \dfrac{ (H_0 - H_{-})(H_0 - H_{+})(-c_1 + R_1) }{ (c_1 + H_0)(c_2 + H_0) } } },
	\end{align}
	\begin{align}
		B_0(H_0)=\frac{\sqrt{c_1 + H_0} \, (c_2 + H_0) \sqrt{H_0 - H_{-}}}{\sqrt{H_0 - H_{+}}},
	\end{align}
	\begin{align}
		B_{\mathrm{F}}(H_0)=\frac{(c_2 + H_0) \sqrt{\dfrac{(c_1 + H_{-})(H_0 - H_{+})}{(c_1 + H_0)(H_{-} - H_{+})}} \; (c_2 + H_{+}) \; (2c_1 + H_{-} + H_{+})}{\sqrt{c_2 - c_1} \; \sqrt{\dfrac{(c_2 + H_0)(c_1 + H_{-})}{(c_1 + H_0)(c_2 + H_{-})}} \; \sqrt{c_2 + H_{-}} \; \sqrt{H_0 - H_{+}}},
	\end{align}
	\begin{align}
		B_{\mathrm{E}}(H_0)=-\frac{ \sqrt{c_1 + H_0} \; \sqrt{c_2 + H_{-}} \; \sqrt{\dfrac{(c_2 + H_0)(H_{-} - H_{+})}{(c_2 + H_{-})(H_0 - H_{+})}} \; \sqrt{c_2 + H_{+}} }{ \sqrt{ \dfrac{ (c_1 + H_0)(H_{-} - H_{+}) }{ (c_1 + H_{-})(H_0 - H_{+}) } } },
	\end{align}
	\begin{align}
		B_{\mathrm{\Pi}}(H_0)=\frac{\sqrt{c_1 + H_{-}} \; (H_0 - H_{+}) \; \sqrt{\dfrac{(c_1 + H_0)(H_{-} - H_{+})}{(c_1 + H_{-})(H_0 - H_{+})}} \; \sqrt{\dfrac{(c_2 + H_0)(H_{-} - H_{+})}{(c_2 + H_{-})(H_0 - H_{+})}} \; (c_1 + c_2 + H_{-} + H_{+})}{\sqrt{c_1 + H_0} \; \sqrt{c_1 + H_{+}}},
	\end{align}
\end{widetext}	

\begin{align}
	\phi(H_0)=\sin^{-1}\left(\frac{\sqrt{H_0 - H_{-}} \sqrt{c_1 + H_{+}}}{\sqrt{c_1 + H_{-}} \sqrt{H_0 - H_{+}}}\right),
\end{align}
\begin{align}
	k=\frac{(c_2 + H_{-})(c_1 + H_{+})}{(c_1 + H_{-})(c_2 + H_{+})},
\end{align}
\begin{align}
	n=\frac{c_1 + H_{-}}{c_1 + H_{+}}.
\end{align}
Our solution contains the combination
\begin{align}
	B_{\mathrm{F}}F(\phi,k)+B_{\mathrm{E}}E(\phi,k)+B_{\mathrm{\Pi}}\Pi(n;\phi,k),
\end{align}
which represents the canonical form of a generic elliptic integral. In general, no further simplification is possible unless a degeneracy occurs or a special parameter choice is imposed \cite{H:1910}. In our case, such a degeneracy arises when $c_1=c_2$, which corresponds to the parameter choice $\epsilon=1$. For this value, the integral simplifies and can be expressed in closed form in terms of elementary functions, as shown in Sec.~\ref{sec:Hamiltonian}.

\section{$\lambda$-Leal function}\label{sec:app:Leal}

Transcendental functions play an indispensable role in theoretical physics, providing the mathematical language necessary to model and describe a vast range of physical phenomena. Their importance is underscored by their presence across all branches of mathematical analysis, a contribution that has been fundamental to the understanding of the universe. In this paper, we are primarily interested in a specific type of such functions known as Leal functions, which have been analyzed in depth in Ref.~\cite{LHN:20}. From this family of functions, we are particularly interested in $\mathrm{L}_{\lambda}\mathrm{sinh}_2 x$ which is found by inverting the following equation:
\begin{align}
	\sinh{y(x)}+y(x)=x.\label{eq:Leal-ordinary}
\end{align}

\begin{figure}[!htbp]
	\centering
	\includegraphics[scale=0.43]{plotLeal.png}
	\caption{\textbf{$\lambda$-Leal function.} The function $\mathrm{L}_{\lambda}\mathrm{sinh}_2x$ is shown as a function of $x$ for various values of the parameter $\lambda$.}\label{fig:Leal}
\end{figure}
This leads to the solution $y(x)=\mathrm{Lsinh}_2x$, which is one function of the Leal family. In the more general case that concerns us in Sec.~\ref{sec:Hamiltonian}, we must invert an equation of the form
\begin{align}
	\sinh{y(x)}+\lambda y(x)=x,\label{eq:Leal-gen}
\end{align}
which clearly reduces to the ordinary Leal function equation \eqref{eq:Leal-ordinary} when $\lambda=1$. Following Ref.~\cite{LHN:20}, we define this function via a series expansion. We start by writing $y(x)$ as a series around $x=0$,
\begin{align}
	y(x)=\sum_{n=0}^{\infty}y_{n}x^n, 
\end{align}
and then, by substituting into Eq.~\eqref{eq:Leal-gen}, we can identify all coefficients $y_n$. To demonstrate how this works, we truncate the series expansion up to $n=8$. First, in order to simplify the relations, we note that $y_0$ must vanish. After this, the polynomial equation we need to solve is
\begin{widetext}
	\begin{align}
		&(-1 + y_1 + \lambda y_1)x + (1 + \lambda)y_2x^2 + \left(\frac{y_1^3}{6} + y_3 + \lambda y_3\right)x^3 + \left(\frac{y_1^2y_2}{2} + y_4 + \lambda y_4\right)x^4 \nonumber \\
		&+ \left(\frac{y_1^5}{120} + \frac{y_1y_2^2}{2} + \frac{y_1^2y_3}{2} + y_5 + \lambda y_5\right)x^5
		+ \left(\frac{y_1^4y_2}{24} + \frac{y_3^2}{6} + y_1y_2y_3 + \frac{y_1^2y_4}{2} + y_6 + \lambda y_6\right)x^6\nonumber \\
		&+ \left(\frac{y_1^7}{5040} + \frac{y_1^3y_2^2}{12} + \frac{y_1^4y_3}{24} + \frac{y_2^2y_3}{2} + \frac{y_1y_3^2}{2} + y_1y_2y_4 + \frac{y_1^2y_5}{2} + y_7 + \lambda y_7\right)x^7 \nonumber \\
		&+ \left(\frac{y_1^6y_2}{720} + \frac{1}{6}y_1^3y_2y_3 + \frac{y_2y_3^2}{2} + \frac{y_1^4y_4}{24} + \frac{y_2^2y_4}{2} + y_1(y_3y_4 + y_2y_5) + \frac{1}{12}y_1^2(y_3^2 + 6y_6) + y_8 + \lambda y_8\right)x^8=0.
	\end{align}
\end{widetext}
Each coefficient of the powers of $x$ must vanish, which yields the results for $y_n$. Consequently, the first terms of the $\lambda$-Leal function are given by
\begin{widetext}
	\begin{align}
		\mathrm{L}_{\lambda}\mathrm{sinh}_2x\equiv y(x)=\frac{1}{1+\lambda}x-\frac{1}{6(1+\lambda)^4}x^3+\frac{9-\lambda}{120(1+\lambda)^7}x^5+\frac{-225+54\lambda-\lambda^2}{5040(1+\lambda)^{10}}x^7+\mathcal{O}(x^9),\label{eq:lambda-Leal-expansion}
	\end{align} 
\end{widetext}
We denote the full series expansion as the $\lambda$-Leal function and write it as $\mathrm{L}_{\lambda}\mathrm{sinh}_2x$. We note that all even powers of $x$ have vanishing coefficients and thus they are absent from the expansion. One major difference from the $\lambda=1$ case, i.e., the ordinary Leal function, is that the sign of $\lambda$ plays a crucial role in the monotonicity of the function, which is a prerequisite for inverting Eq.~\eqref{eq:Leal-gen} in the first place. To analyze this, we use the basic property of inverse functions, $g^{-1}(g(x)) = x$, and differentiate it. For our case we find that
\begin{align}
	\frac{d}{dx}\mathrm{L}_{\lambda}\mathrm{sinh}_2x=\frac{1}{\lambda+\cosh{\left(\mathrm{L}_{\lambda}\mathrm{sinh}_2x\right)}}.
\end{align}
Since $\cosh x \geqslant 1$ for every $x$, monotonicity is guaranteed for $\lambda > -1$. For different values of $\lambda$, the reader may refer to Fig.~\ref{fig:Leal} to gain an idea of how these functions behave.

\section{Matching at $\mathcal R_1=c_\ell$}
\label{sec:app:branch-matching}

We check that the branch surface $\mathcal R_1=c_\ell$ does not
introduce an additional singularity in the inverse map entering
the Lagrangian. Let
\begin{align}
	\Delta H\equiv H_>-H_<,
	\qquad
	u\equiv \sqrt{c_\ell+3H_0}.
\end{align}
Approaching the surface from the low-curvature side,
$\mathcal R_1=c_\ell-\delta$, one finds from the trigonometric
solution
\begin{widetext}
	\begin{align}
		\lambda_{\rm K}
		=
		1+\frac{4\delta}{9c_\ell}+\mathcal{O}(\delta^2),
		\qquad
		\theta
		=
		\frac{2u}{3c_\ell}\sqrt{\delta}
		+\mathcal{O}(\delta^{3/2}),
		\qquad
		\Delta H
		=
		\frac{3c_\ell^2}{\delta}+\mathcal{O}(1).
	\end{align}
\end{widetext}
Similarly, approaching from the high-curvature side,
$\mathcal R_1=c_\ell+\delta$, the Leal/hyperbolic solution gives
\begin{widetext}
	\begin{align}
		\lambda_{\rm L}
		=
		-1+\frac{4\delta}{9c_\ell}+\mathcal{O}(\delta^2),
		\qquad
		Y
		=
		\frac{2u}{3c_\ell}\sqrt{\delta}
		+\mathcal{O}(\delta^{3/2}),
		\qquad
		\Delta H
		=
		\frac{3c_\ell^2}{\delta}+\mathcal{O}(1).
	\end{align}
\end{widetext}
Substituting these expansions into Eqs.~\eqref{eq:kepler-branch} and
\eqref{eq:leal-branch}, one obtains
\begin{align}
	\lambda_{\rm K}\theta-\sin\theta
	=
	\left(
	\frac{8u}{27c_\ell^2}
	+
	\frac{4u^3}{81c_\ell^3}
	\right)\delta^{3/2}
	+\mathcal{O}(\delta^{5/2}),
\end{align}
and
\begin{align}
	\sinh Y+\lambda_{\rm L}Y
	=
	\left(
	\frac{8u}{27c_\ell^2}
	+
	\frac{4u^3}{81c_\ell^3}
	\right)\delta^{3/2}
	+\mathcal{O}(\delta^{5/2}).
\end{align}
Therefore the apparent singular prefactors in
$\mathcal B_{\rm K}$ and $\mathcal B_{\rm L}$ cancel, and both
branches have the common limit
\begin{align}
	\lim_{\mathcal R_1\to c_\ell^-}\mathcal B_{\rm K}
	=
	\lim_{\mathcal R_1\to c_\ell^+}\mathcal B_{\rm L}
	=
	b_0(c_\ell)
	\pm
	\left(
	\frac{2u}{9}
	+
	\frac{u^3}{27c_\ell}
	\right).
\end{align}

The derivative with respect to $\mathcal K_1$ also matches
directly from the branch expressions. Differentiating
$\mathcal B_{\rm K}$ and $\mathcal K_1$ with respect to
$\theta$ gives
\begin{align}
	\partial_{\mathcal K_1}\mathcal B_{\rm K}
	=
	\frac{\Delta H}{4(c_\ell-\mathcal R_1)}
	\frac{(\lambda_{\rm K}-\cos\theta)^3}
	{\lambda_{\rm K}\cos\theta-1}.
\end{align}
Likewise, differentiating $\mathcal B_{\rm L}$ and
$\mathcal K_1$ with respect to $Y$ gives
\begin{align}
	\partial_{\mathcal K_1}\mathcal B_{\rm L}
	=
	\frac{\Delta H}{4(\mathcal R_1-c_\ell)}
	\frac{(\lambda_{\rm L}+\cosh Y)^3}
	{\lambda_{\rm L}\cosh Y+1}.
\end{align}
Using the expansions above, both expressions reduce to
\begin{align}
	\lim_{\mathcal R_1\to c_\ell^\mp}
	\partial_{\mathcal K_1}\mathcal B
	=
	\frac{(2c_\ell+u^2)^3}
	{27c_\ell^2(2c_\ell-u^2)} ,
\end{align}
which is finite away from the usual local-invertibility point
$u^2=2c_\ell$, equivalently $\tilde f'(H_0)=0$.

Finally, expanding the branch expressions one order further at
fixed $\mathcal K_1$ gives the common signed expansion
\begin{align}
	\mathcal B(c_\ell+\rho,\mathcal K_1)
	=
	b_0(c_\ell+\rho)
	\pm
	\left[
	F_0(u)
	+
	\rho F_1(u)
	+
	\mathcal{O}(\rho^2)
	\right],
\end{align}
where
\begin{align}
	\rho=\mathcal R_1-c_\ell, \quad F_0(u)=\frac{2u}{9}+\frac{u^3}{27c_\ell},
\end{align}
and $F_1(u)$ is finite on a nondegenerate inverse branch. Hence
$\partial_{\mathcal R_1}\mathcal B$ has the same one-sided limit
from the trigonometric and Leal branches, assuming
$b_0(\mathcal R_1)$ is smooth.

\section{Four-dimensional curvature scalars}\label{sec:app:curvature-scalars}
In our construction, we have adopted a comoving gauge. While we currently have a spatially covariant Lagrangian, our goal is to upgrade it to the fully covariant $4D$ case. This can be achieved by introducing a scalar reference field, following the standard correspondence between spatially covariant gravity and scalar-tensor theories (see, e.g., Refs.~\cite{FGY:16,GH:20}). The scalar is used to covariantly encode the preferred foliation defined by the deparameterized dust clock. In unitary gauge one sets $\phi=t$, so that the normal vector to the spatial hypersurfaces is identified with the gradient of the scalar field,
\begin{align}
	n_{\mu}=\partial_{\mu}\phi .
\end{align}
To preserve the unit lapse condition $N=1$, the scalar is constrained to be unit timelike,
\begin{align}
	g^{\mu\nu}\partial_{\mu}\phi\partial_{\nu}\phi=-1.
\end{align}
In the four-dimensional action this condition is imposed by a constraint sector, equivalently by the Lagrange multiplier of the mimetic/reference-field formulation or by the dust-clock constraint inherited from the deparameterized description. On the polymerized vacuum branch considered in this work, the corresponding dust density is set to zero. Thus the scalar/reference sector fixes the clock and foliation needed for the covariant completion, but it does not represent an additional matter source or an independent propagating scalar degree of freedom.

Utilizing this construction, we can express the components of the extrinsic curvature of the $t=\mathrm{const}$ slice in terms of derivatives of the scalar field as
\begin{align}
	K_{\mu\nu}=\nabla_{\mu}\nabla_{\nu}\phi.
\end{align}
We denote this operation by $\mathcal{P}$, i.e., $\mathcal{P}(K_{\mu\nu}) = \nabla_{\mu}\nabla_{\nu}\phi$. 

In upgrading to four dimensions, the three-dimensional curvature scalars appearing in the spatially covariant Lagrangian must be rewritten in terms of four-dimensional quantities. We denote the unit normal to the preferred foliation by \(n^a\), with \(n^a n_a=-1\), and the induced metric by
\begin{align}
	h_{ab}=g_{ab}+n_a n_b .
\end{align}
With our sign conventions, the Gauss relation is
\begin{align}
	{}^{(3)}R_{abcd}
	=
	h_a{}^e h_b{}^f h_c{}^g h_d{}^q\,
	{}^{(4)}R_{efgq}
	+K_{ad}K_{bc}-K_{ac}K_{bd}.
	\label{eq:Gauss}
\end{align}
Contracting Eq.~\eqref{eq:Gauss} twice gives
\begin{align}
	{}^{(3)}R
	=
	{}^{(4)}R
	+2{}^{(4)}R_{ab}n^a n^b
	+K_{ab}K^{ab}
	-K^2 .
	\label{eq:3Ricci}
\end{align}
This is the relation used to replace the intrinsic Ricci scalar.

To obtain the corresponding relation for the Kretschmann scalar, we define the fully projected four-dimensional Riemann tensor
\begin{align}
	\mathscr{R}_{abcd}
	\equiv
	h_a{}^e h_b{}^f h_c{}^g h_d{}^q\,
	{}^{(4)}R_{efgq}.
\end{align}
Using Eq.~\eqref{eq:Gauss}, one finds
\begin{widetext}
	\begin{align}
		{}^{(3)}\mathcal{K}
		&\equiv
		{}^{(3)}R_{abcd}{}^{(3)}R^{abcd}
		=
		\mathscr{R}_{abcd}\mathscr{R}^{abcd}
		-4\,{}^{(4)}R_{abcd}K^{ac}K^{bd}
		+2\left[
		\left(K_{ab}K^{ab}\right)^2
		-
		K^a{}_bK^b{}_cK^c{}_dK^d{}_a
		\right].
		\label{eq:3Krets-intermediate}
	\end{align}
\end{widetext}
The remaining projected curvature term can be written entirely in terms of four-dimensional tensors by expanding \(h_{ab}=g_{ab}+n_an_b\):
\begin{widetext}
	\begin{align}
		\mathscr{R}_{abcd}\mathscr{R}^{abcd}
		=
		{}^{(4)}\mathcal{K}
		+4\,{}^{(4)}R_{aebf}\,{}^{(4)}R^{e}{}_{c}{}^{f}{}_{d}
		n^a n^b n^c n^d
		+4\,{}^{(4)}R_{abcd}\,{}^{(4)}R_{e}{}^{bcd}
		n^a n^e .
		\label{eq:projectedRiemann2}
	\end{align}
\end{widetext}
Combining Eqs.~\eqref{eq:3Krets-intermediate} and \eqref{eq:projectedRiemann2}, we obtain
\begin{widetext}
	\begin{align}
		{}^{(3)}\mathcal{K}
		=
		{}^{(4)}\mathcal{K}
		+2\left[
		\left(K_{ab}K^{ab}\right)^2
		-
		K^a{}_bK^b{}_cK^c{}_dK^d{}_a
		\right]
		-4\Big(
		{}^{(4)}R_{abcd}K^{ac}K^{bd}
		-
		{}^{(4)}R_{aebf}\,{}^{(4)}R^{e}{}_{c}{}^{f}{}_{d}
		n^a n^b n^c n^d
		-
		{}^{(4)}R_{abcd}\,{}^{(4)}R_{e}{}^{bcd}
		n^a n^e
		\Big).
		\label{eq:3Krets}
	\end{align}
\end{widetext}

As a simple check of Eq.~\eqref{eq:3Krets}, consider a
constant-curvature foliation,
\begin{align}
	ds^2=-dt^2+a(t)^2\gamma_{ij}dx^idx^j,
	\qquad
	q\equiv \frac{k}{a^2},
	\qquad
	\mathcal H\equiv \frac{\dot a}{a}.
\end{align}
The intrinsic curvature of the spatial slices is
\begin{align}
	{}^{(3)}R_{ijkl}
	=
	q(h_{ik}h_{jl}-h_{il}h_{jk}),
	\qquad
	{}^{(3)}\mathcal K=12q^2 .
\end{align}
On the other hand,
\begin{align}
	K^i{}_j=\mathcal H\delta^i{}_j,
	\qquad
	{}^{(4)}R_{ijkl}
	=
	(\mathcal H^2+q)(h_{ik}h_{jl}-h_{il}h_{jk}).
\end{align}
Substituting these expressions into the right-hand side of
Eq.~\eqref{eq:3Krets}, the projected four-dimensional curvature
term gives \(12(\mathcal H^2+q)^2\), while the extrinsic-curvature
terms give \(12\mathcal H^4\) and the mixed term gives
\(-24\mathcal H^2(\mathcal H^2+q)\). Hence
\begin{align}
	{}^{(3)}\mathcal K
	=
	12(\mathcal H^2+q)^2
	+
	12\mathcal H^4
	-
	24\mathcal H^2(\mathcal H^2+q)
	=
	12q^2,
\end{align}
in agreement with the direct intrinsic computation.

\end{document}